\documentclass[conference]{IEEEtran}
\IEEEoverridecommandlockouts

\usepackage{cite}
\usepackage{amsmath,amssymb,amsfonts}
\usepackage{algorithmic}
\usepackage{graphicx}
\usepackage{textcomp}
\usepackage{xcolor}
\def\BibTeX{{\rm B\kern-.05em{\sc i\kern-.025em b}\kern-.08em
    T\kern-.1667em\lower.7ex\hbox{E}\kern-.125emX}}

\usepackage{amsthm}
\usepackage{booktabs}
\usepackage{algorithm}

\usepackage{array}

\usepackage{subfig}
\usepackage{hyperref}

\title{Robust Knowledge Distillation in Federated Learning: Counteracting Backdoor Attacks%
\thanks{This work has been accepted for publication in the IEEE Conference on Secure and Trustworthy Machine Learning (SaTML). The final version will be available on IEEE Xplore.}}

\author{
    \IEEEauthorblockN{Ebtisaam Alharbi\IEEEauthorrefmark{1}\IEEEauthorrefmark{2}, Leandro Soriano Marcolino\IEEEauthorrefmark{1}, Qiang Ni\IEEEauthorrefmark{1}, and Antonios Gouglidis\IEEEauthorrefmark{1}}
    \IEEEauthorblockA{\IEEEauthorrefmark{1}School of Computing and Communications, Lancaster University, United Kingdom\\
    \IEEEauthorrefmark{2}Department of Computer Science, Umm Al-Qura University, Saudi Arabia\\
    Email: \{e.alharbi, l.marcolino, q.ni, a.gouglidis\}@lancaster.ac.uk}
}

\begin{document}
\maketitle

\begin{abstract}
Federated Learning (FL) enables collaborative model training across multiple devices while preserving data privacy. However, it remains susceptible to backdoor attacks, where malicious participants can compromise the global model. Existing defence methods are limited by strict assumptions on data heterogeneity (Non-Independent and Identically Distributed data) and the proportion of malicious clients, reducing their practicality and effectiveness. To overcome these limitations, we propose Robust Knowledge Distillation (RKD), a novel defence mechanism that enhances model integrity without relying on restrictive assumptions. RKD integrates clustering and model selection techniques to identify and filter out malicious updates, forming a reliable ensemble of models. It then employs knowledge distillation to transfer the collective insights from this ensemble to a global model. Extensive evaluations demonstrate that RKD effectively mitigates backdoor threats while maintaining high model performance, outperforming current state-of-the-art defence methods across various scenarios.
\end{abstract}

\section{Introduction}

Federated Learning (FL) has emerged as a powerful paradigm that enables multiple clients to collaboratively train a shared global model without exchanging their private local data, thereby enhancing data privacy and security \cite{ref_article14}. In FL, each client trains a local model on its own dataset and shares only the model updates with a central server, which aggregates these updates to form the global model. Despite its advantages, FL faces significant challenges that threaten the effectiveness and integrity of the resulting global model.

One critical challenge is the vulnerability of FL to security threats, particularly backdoor attacks. In these attacks, malicious clients introduce harmful updates into the training process by embedding hidden triggers within their model updates \cite{ref_article15}. These hidden triggers are specific patterns or inputs that, when encountered by the global model, cause it to produce incorrect or malicious outputs. Malicious clients achieve this by manipulating their local training data and altering their model parameters, ensuring that the backdoor behaviour is incorporated into the global model during aggregation without being detected. Since these triggers are often designed to be stealthy inputs, and malicious clients train on both clean input data and inputs containing the trigger, the attacks can go unnoticed during normal validation. This stealthiness makes backdoor attacks particularly effective and dangerous threats to FL systems \cite{ref_article25}. As a result, the compromised global model behaves normally on standard inputs but can cause significant harm when the triggers are activated, posing serious risks to the integrity and reliability of FL deployments.

Another significant challenge arises from the Non-IID data distribution across clients. In real-world scenarios, client data often differ substantially from the overall data distribution, leading to inconsistent global model performance and slower convergence \cite{ref_article24}. This heterogeneity can cause conflicting model updates from different clients, making it difficult for the global model to generalize well across all participants. Many existing defense methods assume that client data are Independent and Identically Distributed (IID) and that only a small fraction of clients are malicious \cite{ref_article27},\cite{ref_article28},\cite{ref_article8}. These assumptions simplify the aggregation process but fail to capture the complexity of practical FL deployments.

Various defense strategies have been explored. Robust aggregation methods, such as RLR \cite{ref_article17} and FoolsGold \cite{ref_article8}, modify the aggregation process to reduce the influence of malicious updates; however, they may struggle against sophisticated backdoor attacks and perform suboptimally in Non-IID settings. Other approaches include root data validation methods like FLTrust \cite{ref_article4} and Baffle \cite{ref_article2}, which use trusted data to validate client updates, yet they can be circumvented by adversaries that mimic benign behavior. Differential Privacy-based methods, exemplified by FLAME \cite{ref_article16}, incorporate noise and weight clipping to maintain privacy and integrity, but they require careful tuning and often compromise model performance. While approaches such as FedDF \cite{ref_article13} and FedBE \cite{ref_article5} have been proposed to address the challenges posed by Non-IID data—primarily through knowledge distillation techniques—they focus on improving model performance under data heterogeneity rather than explicitly defending against backdoor attacks. In other words, although these methods help mitigate issues caused by Non-IID distributions, they do not provide targeted defenses against adversarial updates.

To bridge this gap, we propose Robust Knowledge Distillation (RKD), a novel defense mechanism that enhances the integrity of the global model without relying on restrictive IID assumptions or low malicious-client ratios. RKD integrates clustering and model selection techniques to identify and filter out malicious updates, forming a reliable ensemble of benign models. Specifically, it employs cosine similarity measures and Hierarchical Density-Based Spatial Clustering of Applications with Noise (HDBSCAN) to cluster client models, isolating potential outliers. Models closest to the median of the benign cluster are then selected to form an ensemble that is subsequently distilled into the global model. This approach defends against backdoor attacks in Non-IID settings.

Our contributions are summarized as follows:

\begin{itemize}
    \item We introduce the RKD framework, which robustly aggregates client updates by filtering out adversarial contributions without assuming IID data or a low fraction of malicious clients.
    \item We evaluate RKD on diverse datasets, including CIFAR-10, EMNIST, and Fashion-MNIST, demonstrating its effectiveness against sophisticated backdoor attacks such as A3FL \cite{ref_article22}, F3BA \cite{ref_article7}, DBA \cite{ref_article20}, ADBA \cite{ref_article26}, and TSBA \cite{ref_article3}.
    \item Empirical results show that RKD maintains high model accuracy above 80\% while reducing attack success rates below 17\%, outperforming state-of-the-art defense methods, particularly in challenging Non-IID scenarios.
\end{itemize}

Extensive experiments underscore the effectiveness of RKD in enhancing FL robustness, and maintaining high model performance while effectively mitigating backdoor threats. The implementation of the RKD framework is available at \url{https://github.com/EbtisaamCS/RKD.git}.

\section{Background and Problem Setting}

FL is a distributed machine learning paradigm where multiple clients collaboratively train a shared global model \( M_{\text{global}} \) while keeping their local data \( D_i \) decentralized and private. Each client \( i \) trains a local model \( M_i \) on its own dataset \( D_i \) and then shares only the model parameters \( \theta_i \) with a central server. The central server aggregates these local parameters to update the global model. A common aggregation algorithm is Federated Averaging (FedAvg), which computes the global model parameters as: $
\theta_{\text{global}} = \frac{1}{N} \sum_{i=1}^{N} \theta_i,
$ where \( N \) is the number of participating clients \cite{ref_article14}. After aggregation, the server sends the updated global model \( M_{\text{global}} \) back to all participating clients. Each client then updates its local model using the received global parameters, ensuring synchronization across the network. This collaborative process allows the construction of a global model that benefits all clients without exposing datasets, thus preserving data privacy \cite{ref_article25}.

However, despite its advantages, FL is vulnerable to security threats, particularly backdoor attacks. In backdoor attacks, malicious clients aim to compromise the integrity of the global model \( M_{\text{global}} \) by manipulating their local training data \( D_i \) and/or altering their model updates \( \theta_i \) before sending them to the central server. These manipulations are designed to embed hidden malicious behaviours into the global model without being detected. Furthermore, the presence of Non-IID data distributions among clients exacerbates the difficulty in detecting and mitigating such attacks, as data heterogeneity can mask the malicious updates, making it challenging to distinguish between benign model variations and manipulations \cite{ref_article7}.

\subsection{Backdoor Attacks}

\subsubsection{Data Poisoning Attacks}

In backdoor data poisoning attacks, malicious clients intentionally modify their local training data to embed hidden triggers that induce the global model to exhibit malicious behaviour when these triggers are encountered \cite{ref_article15}. Specifically, a malicious client takes its original dataset \( D_i = \{ (x_j, y_j) \} \) and creates a poisoned dataset \( D_i^{\text{poison}} \) by injecting a small number of modified data instances \( \{ (x'_k, y'_k) \} \). In these instances, \( x'_k \) contains a trigger pattern—such as a specific pixel arrangement in an image—and \( y'_k \) is an incorrect target label chosen by the attacker. The modified dataset \( \tilde{D}_i = D_i \cup D_i^{\text{poison}} \) is then used to train the local model \( M_i \) \cite{ref_article20}.

The attacker's objective is to train \( M_i \) such that it performs well on normal data while misclassifying inputs containing the trigger pattern. This can be formalized by minimizing the loss function: $
\mathcal{L}(M_i(\tilde{D}_i), y),
$ where \( y \) represents the labels. By successfully training on \( \tilde{D}_i \), the model learns the backdoor association between the trigger and the target label \( y'_k \). When these malicious model updates \( \theta_i \) are aggregated by the server, the global model \( M_{\text{global}} \) inherits the backdoor behaviour, causing it to misclassify inputs containing the trigger pattern while maintaining normal performance on clean data.

\subsubsection{Model Poisoning Attacks}

In backdoor model poisoning attacks, malicious clients manipulate their local model parameters \( \theta_i \) to embed backdoors into the global model, often in combination with injecting triggers into their training data. Unlike data poisoning attacks that focus primarily on the training data, model poisoning attacks involve directly altering the model parameters to maximize the impact of the backdoor while minimizing the risk of detection \cite{ref_article24}.

In general, malicious clients may:

\begin{itemize}
    \item \textbf{Manipulate Model Parameters}: After training on the poisoned data, attackers further manipulate their model parameters \( \theta_i \) to enhance the backdoor effect or to avoid detection. This can involve:
    \begin{itemize}
        \item \textbf{Adversarial Adaptation}: Adapting the trigger and model parameters to remain effective against changes in the global model during training \cite{ref_article22}.
        
        \item \textbf{Scaling Weights}: Amplifying the impact of the malicious update by scaling the model parameters \cite{ref_article3}:
        $
        \theta_i^{\text{poison}} = \alpha \cdot \theta_i,
        $
        where \( \alpha > 1 \) increases the update's influence on the global model during aggregation.
        
        \item \textbf{Adding Perturbations}: Introducing subtle changes to the model parameters to maintain stealth \cite{ref_article7}:
        $
        \theta_i^{\text{poison}} = \theta_i + \delta,
        $
        where \( \delta \) is a small perturbation designed to embed the backdoor while avoiding detection by defence mechanisms.
    \end{itemize}
\end{itemize}

Advanced optimization techniques, such as Projected Gradient Descent (PGD) \cite{ref_article22}, can be employed to iteratively adjust \( \delta \) to find the optimal perturbation that balances attack success with stealthiness \cite{ref_article19}. In some cases, attackers perform model replacement attacks, where they entirely replace their local model with a maliciously crafted model \( \theta_i^{\text{poison}} \) and scale it to influence the global aggregation \cite{ref_article3} disproportionately. By carefully manipulating both the training data and the model parameters, attackers ensure that their malicious updates blend in with those from benign clients, making the backdoor attack stealthy and difficult to detect.

\subsection{Impact on the Global Model}

When the central server aggregates the model updates, the presence of malicious updates from attackers influences the global model parameters. Specifically, the aggregated global model parameters become: $
\theta_{\text{global}}^{\text{poison}} = \frac{1}{N} \left( \sum_{i \in \mathcal{H}} \theta_i + \sum_{j \in \mathcal{M}} \theta_j^{\text{poison}} \right),
$ where \( \mathcal{H} \) is the set of honest clients and \( \mathcal{M} \) is the set of malicious clients. The malicious updates \( \theta_j^{\text{poison}} \) are designed to inject the backdoor into the global model \( M_{\text{global}}^{\text{poison}} \). As a result, the compromised global model may behaves normally on standard (clean) inputs, maintaining high performance and thus not raising suspicion. However, when presented with inputs containing the backdoor trigger, the model exhibits incorrect or malicious behaviour, such as misclassifying the input or producing outputs desired by the attacker. This stealthy alteration poses significant risks to the integrity and reliability of the FL system, as the backdoor can remain undetected until the trigger is activated.

\subsection{Characteristics of Backdoor Attacks}

Backdoor attacks often exhibit distinctive characteristics in the model updates sent by malicious clients, which can be exploited for detection. Key characteristics include:

\begin{itemize}
    \item \textbf{Angular Deviation}: Malicious model updates may have a different direction in the parameter space compared to updates from benign clients \cite{ref_article3}. This directional difference can be quantified using the cosine similarity (or angular deviation) between the parameter vectors of malicious updates \( \theta_{\text{attack}} \) and benign updates \( \theta_{\text{benign}} \): $
    \Delta \theta_{\text{angular}} = \cos^{-1} \left( \frac{ \theta_{\text{attack}} \cdot \theta_{\text{benign}} }{ \| \theta_{\text{attack}} \| \| \theta_{\text{benign}} \| } \right)$. Angular deviations indicate that the malicious update is contributing in a different direction, which may be due to the attacker's attempt to embed the backdoor, such as F3BA attack \cite{ref_article7}.

    \item \textbf{Magnitude Deviation}: Malicious updates may have a significantly different norm magnitude compared to benign updates. This occurs when attackers scale their model updates to increase their influence on the aggregated global model \cite{ref_article20}. The magnitude deviation can be observed when: $
    \| \theta_{\text{attack}} \| \gg \| \theta_{\text{benign}} \|.
    $ 

    \item \textbf{Subtle Deviations}: Some attackers design their updates to closely resemble those of benign clients, keeping the deviation between the malicious and benign updates within a small threshold \( \epsilon \) to avoid detection: $
    \| \theta_{\text{attack}} - \theta_{\text{benign}} \| < \epsilon.
    $ Techniques like A3FL attack \cite{ref_article22} adjust both the magnitude and direction of the malicious update to embed the backdoor while maintaining stealthiness.

\end{itemize}

Understanding these characteristics is crucial for designing defence mechanisms that can detect and mitigate backdoor attacks without relying on overly restrictive assumptions.

\subsection{Threat Model}

Our work assumes the FL system in which a subset of clients, referred to as \textbf{malicious clients}, aim to compromise the global model through backdoor attacks. The goal of these attacks is to embed hidden triggers into the global model, causing targeted misclassifications on specific inputs while maintaining high accuracy on clean data to avoid detection. Below, we outline the adversary's knowledge, capabilities, and the defender's knowledge:

\begin{itemize}
    \item \textbf{Adversary's Knowledge:} 
    The malicious clients have no access to other clients' model updates or data. They operate under the assumption that they can only manipulate their local data and model updates.
    
    \item \textbf{Adversary's Capabilities:} 
    The adversaries can collude and coordinate their attacks but cannot intercept or alter communications between other clients and the server. They can manipulate their local training data (\textit{data poisoning}) and modify their model updates before sending them to the server (\textit{model poisoning}).
    
    \item \textbf{Defender's Knowledge:} 
    The server is aware that some clients may be malicious but does not know their identities. The server has access only to the submitted model updates and any auxiliary data (e.g., public unlabeled datasets) used for knowledge distillation.
\end{itemize}

\section{Related Work}

Defending against backdoor attacks in FL is a critical area of research, with various strategies proposed to enhance the robustness of FL. These defence mechanisms can be broadly categorized into robust aggregation methods, clustering-based defences, and knowledge distillation approaches.

\textbf{Robust Aggregation Methods} aim to mitigate the influence of malicious updates during the model aggregation process. The Robust Learning Rate (RLR) method \cite{ref_article17} adjusts the learning rates of clients based on the alignment of their updates with the global model's direction. By assigning smaller learning rates to updates that deviate significantly, RLR reduces the impact of potentially malicious updates. However, RLR may not fully address attacks that manipulate the magnitude of updates, such as scaling attacks, and its effectiveness diminishes in heterogeneous (Non-IID) data environments where benign updates naturally vary in direction and magnitude. 

FoolsGold \cite{ref_article8} is designed to counter backdoor attacks by analyzing the similarity of gradient updates among clients. It assigns lower aggregation weights to clients whose updates are overly similar, under the assumption that malicious clients will produce highly similar gradients due to coordinated attacks. While effective against certain types of attacks, FoolsGold may inadvertently penalize benign clients with similar data distributions, leading to unfairness and potential degradation of overall model performance.

\textbf{Clustering-Based Defences.} Prior work has explored clustering-based techniques to distinguish between benign and malicious updates in federated learning by grouping similar updates and flagging outliers. For example, FLAME \cite{ref_article16} leverages HDBSCAN for clustering and introduces noise to enhance security; however, its adaptability to evolving threat landscapes remains limited. Similarly, RFCL \cite{ref_article1} employs HDBSCAN to cluster client updates for defending against gradient poisoning attacks, yet its effectiveness diminishes against adaptive backdoor attacks \cite{ref_article22}. A critical limitation of these methods lies in their reliance on the precision of the clustering process, which is often compromised in high-dimensional parameter spaces typical of deep learning models. In such spaces, traditional clustering algorithms can struggle due to the curse of dimensionality, leading to inefficient clustering and the potential misclassification of benign updates as malicious, or vice versa.

In contrast, our approach mitigates these issues by first computing cosine similarity scores between client updates and the global model. These scalar similarity values capture the directional alignment of updates while reducing the data from a high-dimensional space to a one-dimensional representation, thereby simplifying the clustering task. We then apply HDBSCAN to these cosine similarity scores, which significantly enhances clustering efficiency and accuracy. 

\textbf{Knowledge Distillation Approaches} have been recognized for their ability to improve learning efficiency and address Non-IID data distributions in a federated framework. Methods such as FedDF \cite{ref_article13} and FedBE \cite{ref_article5} leverage knowledge distillation to aggregate models. FedDF performs model fusion by distilling knowledge from client models into a global model using unlabeled public data, effectively aligning the models’ outputs. FedBE builds upon this by employing Bayesian ensemble techniques to instruct a student model based on the ensemble’s predictive distribution.

FedRAD \cite{ref_article18} enhances FedDF by assigning weights to client models according to their median scores, which measure how often a model's prediction corresponds to the median prediction among all clients. While these methods improve performance in Non-IID settings, they are vulnerable to backdoor attacks when transferring knowledge from ensemble models without analyzing outliers. Malicious clients can introduce backdoors into their models, and without mechanisms to detect and exclude these compromised models, the backdoor triggers can be propagated to the global model during distillation.

Our proposed approach, \textbf{RKD}, builds upon existing methods by innovatively integrating and enhancing foundational techniques such as HDBSCAN and knowledge distillation. This approach addresses the limitations of prior works, particularly in defending against recent backdoor attacks in Non-IID data.

First, by integrating cosine similarity with HDBSCAN, RKD improves clustering efficiency in high-dimensional parameter spaces, effectively identifying malicious updates even when they are subtle or adaptive. Focusing on the angular relationships between model updates captures essential differences without being overwhelmed by the volume of parameters, thereby ensuring accurate clustering despite Non-IID data.

Second, the RKD framework incorporates a robust model selection process that selects models near the median of the cluster. By identifying and using the most representative models, RKD further mitigates the influence of any remaining malicious updates. This selection forms a reliable and trustworthy ensemble for knowledge distillation.

Third, RKD ensures secure knowledge transfer by distilling from the carefully selected ensemble models, preventing the propagation of backdoor triggers to the global model. By excluding outlier models identified during clustering, RKD reduces the risk of incorporating malicious behaviours into the model. Additionally, knowledge distillation aids in smoothing out variations caused by Non-IID data, leading to a more generalized and robust global model.

By addressing the challenges of high-dimensional data, improving clustering efficiency, and ensuring secure knowledge transfer, RKD provides a robust defence against backdoor attacks in federated learning while effectively handling Non-IID data distributions.

\section{Methodology}

The RKD framework is designed to secure FL against backdoor attacks by identifying and mitigating malicious model updates. It consists of three core components: \textit{Automated Clustering}, \textit{Model Selection}, and a \textit{Knowledge Distillation Module}. These components work together to detect and eliminate malicious influences while preserving the performance and integrity of the global model.

\subsection{Framework Overview}
The proposed RKD framework employs a multi-tiered strategy to enhance the robustness of federated learning systems. We provide an overview of the framework by outlining its key components and processes. Detailed explanations of each component are presented in the subsequent subsections.

Initially, the central server initializes the global model \( M_{\text{global}}^0 \) and broadcasts it to all participating clients. Each client \( i \) trains its local model \( M_i^r \) on its private dataset \( D_i \), starting from the current global model \( M_{\text{global}}^r \). The resulting updated model parameters \( \theta_i^r \) are then returned to the server.

At the server, the focus is on identifying potential malicious updates. The server computes the cosine similarity between each client model update \( \theta_i^r \) and the current global model \( M_{\text{global}}^r \). This metric captures the angular similarity between local updates and the global model, enabling the detection of updates that significantly deviate in direction—a characteristic of malicious behavior known as \textit{Angular Deviation}.

Using these similarity scores, the server employs the HDBSCAN algorithm to cluster the models based on their similarity to the global model. This clustering process distinguishes benign updates, which form dense clusters due to their similarity, from potentially malicious updates, which appear as outliers because of their dissimilarity. By clustering based on similarity scores rather than directly using high-dimensional parameter vectors, the computational complexity is significantly reduced, making the clustering process more scalable and efficient.

Within the benign cluster, the server computes the median of the model parameters by taking the median of each parameter across the models. This helps mitigate the impact of extreme values in the updates, addressing the \textit{Magnitude Deviation} characteristic of malicious behavior.

To further enhance robustness, the server selects models whose parameters are closest to the median to form an ensemble. This selection filters out any updates that may have evaded detection during clustering but still deviate from the central tendency of the benign models.

The selected models are then aggregated to form an initial distilled model. To further refine this model and improve its resilience against subtle backdoor triggers in Non-IID settings, Knowledge Distillation (KD) is applied. The ensemble of selected models guides the refinement of the distilled model \( M_{\text{global}}^{r+1} \), ensuring it reflects the collective behavior of benign clients. This process mitigates the risk of \textit{Subtle Deviations}, where attackers craft updates that closely mimic benign updates in both magnitude and direction.

Finally, the server broadcasts the refined global model \( M_{\text{global}}^{r+1} \) to benign clients for further training. For clients identified as malicious, our primary approach is the \textbf{Exclusion Strategy}: these clients are not provided with the updated global model \( M_{\text{global}}^{r+1} \); instead, they continue training using their current local model until they are reclassified as benign. This complete exclusion prevents adversaries from adapting their strategies based on the most recent global model updates.

As an alternative, we also consider the \textbf{Perturbation Strategy} (referred to as RKD Perturbed Global Model, or PGM). In this approach, to mask the fact that a client has been flagged as malicious, the server sends a perturbed version of the refined global model. Specifically, the perturbed model is defined as
\(
M_{\text{pert}}^{r+1} = M_{\text{global}}^{r+1} + \eta,
\)
where \(\eta\) is a noise vector with a small magnitude (e.g., \(\|\eta\| \approx 1 \times 10^{-4}\)). This strategy limits the opportunity for adaptive adversaries to deduce their status while still providing them with an updated model.

In both cases, if a client previously classified as malicious is reclassified as benign in subsequent rounds, it will begin receiving the updated global model \( M_{\text{global}}^{r+1} \) to realign with the overall training process. 

Algorithm~\ref{alg:RKD} summarizes the RKD framework.

\begin{algorithm}[htp]
\caption{RKD Framework Methodology} \label{alg:RKD}
\begin{algorithmic}[1]
\REQUIRE Clients \( \mathcal{A} \), number of iterations \( R \), malicious client strategy \( S \in \{\text{Exclusion, Perturbation}\} \)
\ENSURE Final global model \( M_{\text{global}}^R \)
\STATE Initialize global model \( M_{\text{global}}^0 \)
\FOR{\( r = 0 \) to \( R-1 \)}
    \IF{\( r = 0 \)}
        \STATE Send \( M_{\text{global}}^0 \) to all clients in \( \mathcal{A} \)
    \ELSE
        \STATE Send \( M_{\text{global}}^r \) to benign clients \( \mathcal{A}_{\text{benign}}^{r-1} \)
        \IF{\( S = \text{Exclusion} \)}
            \FOR{each malicious client \( i \in \mathcal{A} \setminus \mathcal{A}_{\text{benign}}^{r-1} \)}
                \STATE Send the current local model \( M_i^r \) to client \( i \)
            \ENDFOR
        \ELSE[Otherwise, using Perturbation]
            \STATE Compute perturbed model \( M_{\text{pert}}^r = M_{\text{global}}^r + \eta \)
            \STATE Send \( M_{\text{pert}}^r \) to malicious clients \( \mathcal{A} \setminus \mathcal{A}_{\text{benign}}^{r-1} \)
        \ENDIF
    \ENDIF
    \STATE Collect models \( \{ M^r \} = \{ M_i^r \mid i \in \mathcal{A} \} \)
    \STATE Identify \( \{ M_{\text{benign}}^r \} \) and \( \mathcal{A}_{\text{benign}}^r \) via clustering \hfill \(\triangleright\) See Algorithm \ref{alg:DM}
    \STATE Select ensemble models \( \mathcal{E}^r \) from \( \{ M_{\text{benign}}^r \} \)
    \STATE Compute aggregated model \( M_{\text{distill}}^r \) from \( \mathcal{E}^r \)
    \STATE Update 
    \(
    M_{\text{global}}^{r+1}= \text{KD}(M_{\text{distill}}^r, \mathcal{E}^r)
    \)
    \hfill \(\triangleright\) See Algorithm \ref{alg:KD}
\ENDFOR
\RETURN Final global model \( M_{\text{global}}^R \)
\end{algorithmic}
\end{algorithm}

\subsection{Automated Clustering.} This component identifies and excludes potentially malicious models to maintain the integrity of the federated learning process. The method leverages cosine similarity and the HDBSCAN clustering algorithm to distinguish between benign and malicious model updates.

At iteration \( r \), each client \( i \) has local model parameters \( \theta_i^r \). To identify alignment with the current global model, the server computes the cosine similarity \( s_i \) between each client’s model and the global model \( M_{\text{global}}^r \):
\[
s_i = \frac{\left( \theta_i^r \right)^\top M_{\text{global}}^r}{\left\| \theta_i^r \right\| \left\| M_{\text{global}}^r \right\|}, \quad \text{for } i = 1, \ldots, N.
\]
Higher cosine similarity values indicate greater alignment with the global model, which is expected for benign clients, while malicious clients are more likely to deviate.

These similarity scores \( \{ s_i \} \) are then clustered using HDBSCAN. Clustering based on scalar similarity scores significantly reduces computational complexity compared to clustering in the high-dimensional parameter space.

The minimum cluster size \( Q \) for HDBSCAN is dynamically adjusted at each training round \( r \) using the formula:
\(
Q = \max\left(2, \left\lceil 0.2N - r \right\rceil\right),
\)
where \( N \) is the total number of participating clients, and \( \lceil \cdot \rceil \) denotes the ceiling function to ensure \( Q \) is an integer.

Applying HDBSCAN to the set of similarity scores yields cluster labels \( \{ L_i \} \):
\(
\{ L_i \} = \text{HDBSCAN}\left( \{ s_i \}, Q \right).
\)
For each identified cluster \( C_k \), the mean cosine similarity \( \mu_k \) is computed:
\(
\mu_k = \frac{1}{\left\| C_k \right\|} \sum_{i \in C_k} s_i,
\)
where \( \left\| C_k \right\| \) is the number of clients in cluster \( C_k \).

The cluster with the highest mean cosine similarity \( \mu_{\text{max}} \) is considered the benign cluster:
\(
\mu_{\text{max}} = \max\left( \{ \mu_k \} \right).
\)
Clusters are classified as follows:
\[
\text{Cluster } C_k \text{ is classified as }
\begin{cases}
\text{benign}, & \text{if } \mu_k = \mu_{\text{max}}, \\
\text{malicious}, & \text{otherwise}.
\end{cases}
\]
Models in the benign cluster are used for further training and receive the updated global model in the next iteration. For models identified as malicious, our framework supports two strategies. In the primary \textbf{Exclusion Strategy}, malicious clients are entirely excluded from receiving the updated global model, thereby limiting their influence on the federated learning process. Alternatively, under the \textbf{Perturbation Strategy}, malicious clients receive a perturbed version of the updated global model, which similarly limits their influence while obscuring their classification. 

The detailed steps of the automated clustering procedure are presented in Algorithm~\ref{alg:DM}.

\begin{algorithm}[htp]
\caption{Automated Clustering Algorithm}\label{alg:DM}
\begin{algorithmic}[1]
\REQUIRE \( \{ M^r \} = \{ \theta_i^r \}_{i=1}^N \): Client models at iteration \( r \)
\ENSURE \( \{ M_{\text{benign}}^r \} \), \( \mathcal{A}_{\text{benign}}^r \): Benign models and client indices

\FOR{each client \( i = 1 \) to \( N \)}
    \STATE Compute cosine similarity score
    \(
    s_i = \frac{\left( \theta_i^r \right)^\top M_{\text{global}}^r}{\left\| \theta_i^r \right\| \left\| M_{\text{global}}^r \right\|}
    \)
\ENDFOR

\STATE Apply HDBSCAN to similarity scores

\FOR{each cluster \( C_k \)}
    \STATE Compute mean cosine similarity
    \(
    \mu_k = \frac{1}{\left\| C_k \right\|} \sum_{i \in C_k} s_i
    \)
\ENDFOR

\STATE Determine benign cluster:
\(
\mu_{\text{max}} = \max\left( \{ \mu_k \} \right)
\)

\STATE Identify benign clients:
\(
\mathcal{A}_{\text{benign}}^r = \left\{ i \mid L_i = k \text{ and } \mu_k = \mu_{\text{max}} \right\}
\)

\STATE Collect benign models:
\(
\{ M_{\text{benign}}^r \} = \left\{ \theta_i^r \mid i \in \mathcal{A}_{\text{benign}}^r \right\}
\)

\RETURN \( \{ M_{\text{benign}}^r \} \), \( \mathcal{A}_{\text{benign}}^r \)
\end{algorithmic}
\end{algorithm}

\subsection{Model Selection.}
This component refines the set of benign models by selecting the most representative ones for aggregation, thereby mitigating the impact of outliers and enhancing the robustness of the aggregated model against backdoor attacks.

From the set of benign models 
\(
\{ M_{\text{benign}}^r \} = \{ \theta_i^r \mid i \in \mathcal{A}_{\text{benign}}^r \},
\)
we compute the median model parameter vector \( \theta_{\text{median}}^r \) by taking the element-wise median across all benign models:
\(
\theta_{\text{median}}^r = \text{median}\left( \{ \theta_i^r \mid i \in \mathcal{A}_{\text{benign}}^r \} \right).
\)

Next, we calculate the distance between each benign model \( \theta_i^r \) and the median model \( \theta_{\text{median}}^r \) using the \( L_1 \) norm:
\(
d_i = \left\| \theta_i^r - \theta_{\text{median}}^r \right\|_1, \quad \text{for all } i \in \mathcal{A}_{\text{benign}}^r.
\)

These distances \( \{ d_i \} \) are then used to rank the benign models. To select the most representative models, we define a threshold \( \epsilon \) based on the dispersion of the \( d_i \) values. For example, one may set
\(
\epsilon = \mu_d + k \sigma_d,
\)
where \( \mu_d \) and \( \sigma_d \) denote the mean and standard deviation of the distances \( \{ d_i \} \), and \( k \) is a hyperparameter that controls the threshold level. The ensemble is then defined as:
\(
\mathcal{E}^r = \left\{ \theta_i^r \in \{ M_{\text{benign}}^r \} \mid d_i \leq \epsilon \right\}.
\)
This threshold ensures that only models whose distances from the median are within a reasonable range are selected, thereby accommodating the inherent variability in benign updates.

\subsection{Knowledge Distillation Process.}
This process refines the global model by distilling knowledge from an ensemble of selected benign models \( \mathcal{E}^r \), identified through the previous steps. The server utilizes an unlabeled dataset \( D_{\text{val}} \) (which is separate from the clients' training data and comprises 16\% of the total training data) for knowledge distillation. Pseudo-labels are generated from the ensemble's outputs to guide the training of the distilled model.

For each sample \( x \) in \( D_{\text{val}} \), the server computes the \textit{logits} using each model in the ensemble. These logits are averaged to produce the ensemble logits:
\[
\text{Ensemble\_Logits}(x) = \frac{1}{\left\| \mathcal{E}^r\right\|} \sum_{M_i \in \mathcal{E}^r} f_{M_i}(x),
\]
where \( f_{M_i}(x) \) denotes the logits (i.e., the raw output scores before softmax) produced by model \( M_i \) for input \( x \). The pseudo-labels are then generated by applying the softmax function with a temperature parameter \( T \):
\[
\tilde{y}(x) = \text{softmax}\left( \frac{\text{Ensemble\_Logits}(x)}{T} \right).
\]
Dividing by \(T\) smooths the probability distribution, allowing for a softer target during distillation.

The distilled model \( M_{\text{distill}} \) is trained to minimize the Kullback-Leibler (KL) divergence between its output probabilities and the pseudo-labels:
\[
\mathcal{L} = D_{\text{KL}}\left( \tilde{y}(x) \, \Big\| \, \text{softmax}\left( \frac{f_{M_{\text{distill}}}(x)}{T} \right) \right).
\]
This loss encourages the distilled model to align with the behavior of the ensemble.

To further stabilize training and enhance generalization, we employ \textit{Stochastic Weight Averaging (SWA)}. SWA maintains a running average of the model weights, capturing the trajectory of the parameters as they converge. During each epoch of knowledge distillation, after updating \( M_{\text{distill}} \) using gradient descent, the SWA model \( M_{\text{SWA}} \) is updated as follows:
\[
M_{\text{SWA}} \leftarrow \frac{n_{\text{SWA}} \cdot M_{\text{SWA}} + M_{\text{distill}} }{ n_{\text{SWA}} + 1 },
\]
where \( n_{\text{SWA}} \) is the number of SWA updates. Initially, \( M_{\text{SWA}} \) is set to \( M_{\text{distill}} \) and \( n_{\text{SWA}} = 1 \).

Finally, after completing the distillation process, the SWA model is set as the updated global model for the next iteration:
\[
M_{\text{global}}^{r+1} \leftarrow M_{\text{SWA}}.
\]

The process of knowledge distillation is summarized in Algorithm~\ref{alg:KD}.

\begin{algorithm}[htp]
\caption{Knowledge Distillation Process} \label{alg:KD}
\begin{algorithmic}[1]
\REQUIRE \( \mathcal{E}^r \): Ensemble of selected benign models; \( D_{\text{val}} \): Unlabeled data for distillation; \( T \): Temperature for softmax; \( E_{\text{KD}} \): Number of epochs; \( \eta \): Learning rate
\ENSURE Updated global model \( M_{\text{global}}^{r+1} \)
\STATE 
\(
\text{Ensemble\_Logits}(x) = \frac{1}{\|\mathcal{E}^r\|} \sum_{M_i \in \mathcal{E}^r} f_{M_i}(x)
\hfill \triangleright \text{ Compute ensemble logits for all } x \in D_{\text{val}}
\)

\STATE Generate pseudo-labels:
\[
\tilde{y}(x) = \text{softmax}\left( \frac{\text{Ensemble\_Logits}(x)}{T} \right)
\]
\STATE Initialize \( M_{\text{SWA}} \leftarrow M_{\text{distill}} \) and \( n_{\text{SWA}} \leftarrow 1 \)
\FOR{epoch \( e = 1 \) to \( E_{\text{KD}} \)}
    \FOR{each mini-batch \( \{ x_b \} \subset D_{\text{val}} \)}
        \STATE 
\(
\text{Distill\_Logits}(x_b) = f_{M_{\text{distill}}}(x_b)
\hfill \triangleright \text{ Compute distilled model logits.}
\)
        \STATE Compute loss:
        \(
        \mathcal{L} = D_{\text{KL}}\left( \tilde{y}(x_b) \, \Big\| \, \text{softmax}\left( \frac{\text{Distill\_Logits}(x_b)}{T} \right) \right)
        \)
        \STATE Update \( M_{\text{distill}} \) using SGD with learning rate \( \eta \)
    \ENDFOR
    \STATE Update SWA model:
    \(
    M_{\text{SWA}} \leftarrow \frac{n_{\text{SWA}} \cdot M_{\text{SWA}} + M_{\text{distill}} }{ n_{\text{SWA}} + 1 }
    \)
    \STATE \( n_{\text{SWA}} \leftarrow n_{\text{SWA}} + 1 \)
\ENDFOR
\STATE Set \( M_{\text{global}}^{r+1} \leftarrow M_{\text{SWA}} \)
\RETURN Updated global model \( M_{\text{global}}^{r+1} \)
\end{algorithmic}
\end{algorithm}

\section{Experiments}

We evaluate the effectiveness of the proposed RKD framework in the FL setting under backdoor attack scenarios. We simulate a standard FL environment where multiple clients collaboratively train a global model under the coordination of a central server. The training process is iterative and continues until convergence is achieved.

\subsection{Datasets and Models}

We conducted experiments on three widely used datasets: CIFAR-10, EMNIST, and Fashion-MNIST:

\textbf{CIFAR-10} \cite{ref_article12} consists of 60,000 color images of size $32 \times 32$ pixels, categorized into 10 classes. It serves as a comprehensive benchmark for image classification tasks.

\textbf{EMNIST} \cite{ref_article6} is an extension of the MNIST dataset, containing $814,255$ handwritten character images across $62$ classes, including digits and letters. The images are grayscale with a $28 \times 28$ pixels resolution.

\textbf{Fashion-MNIST} \cite{ref_article23} comprises $70,000$ grayscale images of fashion products from $10$ categories, each of size $28 \times 28$ pixels, providing a more challenging alternative to MNIST dataset.

For each dataset, we employed model architectures suitable for the complexity of the tasks:

For \textbf{CIFAR-10}, we used a ResNet-18 architecture \cite{ref_article9}, which is well-suited for handling the complexity of colour images and capturing hierarchical features. The model was trained with a batch size of 64 and an initial learning rate of 0.01.

For \textbf{EMNIST}, we utilized a Convolutional Neural Network (CNN) consisting of two convolutional layers, each followed by max pooling and dropout layers to prevent overfitting, and a fully connected layer for classification. The model was trained with a batch size of 64 and a learning rate of 0.001.

For \textbf{Fashion-MNIST}, we implemented a CNN with two convolutional layers, each followed by batch normalization and dropout layers, and a fully connected layer. This architecture aids in normalizing the input features and regularizing the model. The training was conducted with a batch size of 64 and a learning rate of 0.001.
\subsection{Attack Setup}
To evaluate the robustness of the RKD framework against backdoor attacks, we simulated the FL environment with \(30\) clients. We considered three scenarios where \(20\%\), \(40\%\), and \(60\%\) of the clients were compromised by an adversary.

Each compromised client injected backdoor triggers into \(50\%\) of its local training data. The backdoor trigger was a pattern added to the images, and the labels of the poisoned samples were altered to a target class specified by the adversary. This simulates a realistic attack where malicious clients attempt to implant a backdoor into the global model while maintaining normal performance on clean data.

The compromised clients followed the federated learning protocol but aimed to influence the global model towards the backdoor task. The benign clients trained on their local data without any manipulation.

\subsection{Attack Methods}

To thoroughly evaluate the robustness of our RKD framework, we examined its efficacy against four recent and sophisticated backdoor attack methods. These attacks are designed to circumvent traditional defence mechanisms and present significant challenges in federated learning environments.

\paragraph{Adversarially Adaptive Backdoor Attack to Federated Learning (A3FL)} The A3FL attack \cite{ref_article22} enhances backdoor persistence by dynamically adapting the trigger pattern in response to the global training dynamics. Instead of using a static trigger, A3FL continuously optimizes the trigger to remain effective against the evolving global model. It employs an adversarial adaptation loss function and Projected Gradient Descent (PGD) to refine the backdoor trigger. By iteratively updating the trigger based on both the current global model and adversarially crafted models, A3FL ensures that the backdoor remains functional throughout multiple training rounds, making it particularly challenging to detect and mitigate.

\paragraph{Focused-Flip Federated Backdoor Attack (F3BA)} F3BA \cite{ref_article7} targets the global model by manipulating a small subset of its parameters with minimal impact on overall performance. The attack calculates the importance of each parameter \( w[j] \) concerning the global loss \( L_g \) using the following metric: $
S[j] = -\left( \frac{\partial L_g}{\partial w[j]} \right) \cdot w[j]
$. Here, \( \frac{\partial L_g}{\partial w[j]} \) is the gradient of the global loss with respect to parameter \( w[j] \), and \( \cdot \) denotes element-wise multiplication. 

Parameters with the highest importance scores are selected, and their signs are flipped to embed the backdoor. This selective manipulation allows the attacker to implant the backdoor with minimal deviation from normal training behaviour, thereby evading detection.

\paragraph{Distributed Backdoor Attack (DBA)} The DBA \cite{ref_article20} spreads a trigger pattern across multiple adversarial clients, enhancing stealth and making detection more difficult. Each compromised client injects a portion of the full trigger into its local training data. When these local models are aggregated, the global model inadvertently learns to associate the combined trigger pattern with the target class. As a result, inputs containing the full trigger pattern are misclassified, effectively executing the backdoor attack without any single client contributing a suspiciously large modification.

\paragraph{The Anti-Distillation Backdoor Attack (ADBA)} The ADBA \cite{ref_article26}, initially designed for centralized machine learning scenarios utilizing knowledge distillation, has been adapted in this work to the federated learning paradigm. Originally, ADBA embeds a robust backdoor into a teacher model by dynamically optimizing a trigger and using a shadow model to ensure backdoor transfer through distillation. In extending this attack to FL, the global model is treated as the teacher, while client models act as distributed learners, effectively mirroring the hierarchical structure of knowledge distillation. Compromised clients inject backdoors by optimizing a crafted trigger using the PGD while adhering to FL constraints. 
\paragraph{Train-and-Scale Backdoor Attack (TSBA)} TSBA \cite{ref_article3} aims to evade anomaly detection mechanisms by carefully scaling the model weights after training. The adversarial client first trains its local model with the backdoor trigger until convergence. Then, it scales down the model updates by a factor \(\gamma\) that remains within acceptable bounds defined by the FL protocol. By doing so, the malicious updates appear similar in magnitude to benign ones, thereby bypassing defenses that rely on detecting abnormal update sizes. The scaling factor \(\gamma\) is chosen to balance stealth and backdoor effectiveness.

\subsection{Heterogeneous Setting}

In FL, data heterogeneity (Non-IID data distribution) among clients is a common and challenging issue that can significantly impact the learning process and model performance. To simulate realistic federated learning environments with varying degrees of data heterogeneity, we utilize the Dirichlet distribution to partition the datasets among the clients \cite{ref_article22}.

The Dirichlet distribution allows us to control the level of heterogeneity by adjusting the concentration parameter \( \alpha \). Each client \( i \) receives a proportion \( p_{i,k} \) of data from class \( k \), where \( p_i = [p_{i,1}, p_{i,2}, \ldots, p_{i,C}] \sim \text{Dirichlet}(\alpha) \) and \( C \) is the number of classes. A lower \( \alpha \) value leads to more uneven class distributions among clients, simulating higher heterogeneity.

Specifically, we set the \( \alpha \) parameter to:

\textbf{Extreme Heterogeneity:} \( \alpha = 0.5 \), \( \alpha = 0.3 \), and \( \alpha = 0.1 \). These lower values of \( \alpha \) result in clients having data predominantly from a few classes, leading to highly Non-IID data distribution scenarios.

\textbf{Moderate Heterogeneity:} \( \alpha = 0.9 \) and \( \alpha = 0.7 \). Higher values of \( \alpha \) produce more balanced class distributions across clients, representing a moderately heterogeneous setting.

\textbf{IID:} In addition to the above Non-IID scenarios, we conduct experiments under IID conditions (i.e., when data is independently and identically distributed across clients) in the \textbf{appendix} to provide a baseline for comparison.

By varying \( \alpha \) (and including IID experiments), we create a spectrum of data heterogeneity scenarios to comprehensively evaluate the robustness of the RKD framework under different degrees of Non-IID data distributions. This approach allows us to assess how well the RKD framework can handle the challenges posed by data heterogeneity, which is critical for practical federated learning applications.

\textbf{Compared Defence Baselines.} We evaluate the RKD framework against seven recent FL defences methods: \emph {FedAvg} \cite{ref_article14}, \emph{FLAME} \cite{ref_article16}, \emph{FedDF} \cite{ref_article13}, \emph{FedRAD} \cite{ref_article18}, \emph{FedBE} \cite{ref_article5}, \emph{RLR} \cite{ref_article17}, and \emph{FoolsGold (FG)} \cite{ref_article8}. 

\subsection{Evaluation Metrics}

We utilized two key evaluation metrics: \textit{Main Task Accuracy (MTA)} and \textit{Attack Success Rate (ASR)}. These metrics provide a comprehensive understanding of the model's performance on legitimate tasks and its resistance to backdoor triggers.

\paragraph{Main Task Accuracy (MTA)} 
MTA measures the classification accuracy of the global model on a clean test dataset \(D_{\text{test}}\), reflecting its ability to correctly predict the true labels of inputs without any backdoor influence. It is defined as:
\(
\text{MTA} = \frac{\| \{ x \in D_{\text{test}} \mid f(x) = y \} \|}{\|D_{\text{test}}\|},
\)
where \(x\) is an input sample from \(D_{\text{test}}\), \(f(x)\) denotes the prediction of the global model \(f\) for input \(x\), and \(y\) is the corresponding true label. Here, \(\|\cdot\|\) denotes the cardinality of the set.

A higher MTA indicates that the model performs well on the primary classification task, correctly identifying inputs as per their true labels.

\paragraph{Attack Success Rate (ASR)}
ASR evaluates the effectiveness of the backdoor attack by measuring the proportion of poisoned inputs that the global model misclassifies into the attacker's target class. It is calculated on a test dataset containing backdoor triggers, denoted as \(D_{\text{poison}}\):
\(
\text{ASR} = \frac{\left\|\{ x \in D_{\text{poison}} \mid f(x) = y_{\text{target}} \}\right\|}{\left\|D_{\text{poison}}\right\|},
\)
where \(x\) is an input sample from the poisoned test dataset \(D_{\text{poison}}\) that contains backdoor triggers, \(f(x)\) is the prediction of the global model \(f\) for input \(x\), and \(y_{\text{target}}\) is the target label that the attacker intends the model to output for backdoor inputs.

A lower ASR signifies better robustness against backdoor attacks, as it indicates that the model is less likely to misclassify backdoor inputs into the attacker's target class.

The goal of an effective Defence mechanism like the RKD framework is to maintain a high MTA while minimizing the ASR. This balance ensures that the model retains its performance on legitimate data while being resilient to manipulation attempts by adversaries. In our experiments, we focus on achieving this balance to demonstrate the RKD framework's capability to defend against sophisticated backdoor attacks without degrading the overall model performance.

\subsection{Experimental Results}

We evaluated the robustness of the RKD framework against advanced backdoor attacks in FL. The models were trained under non-IID data distributions, measuring the MTA and ASR. To ensure reliability, all experiments were repeated five times with different data resampling, with confidence intervals reported at a significance level of \( \rho = 0.01 \).

\begin{figure}[h!]
\centering
{\includegraphics[width=0.325\columnwidth,height=2.5cm]{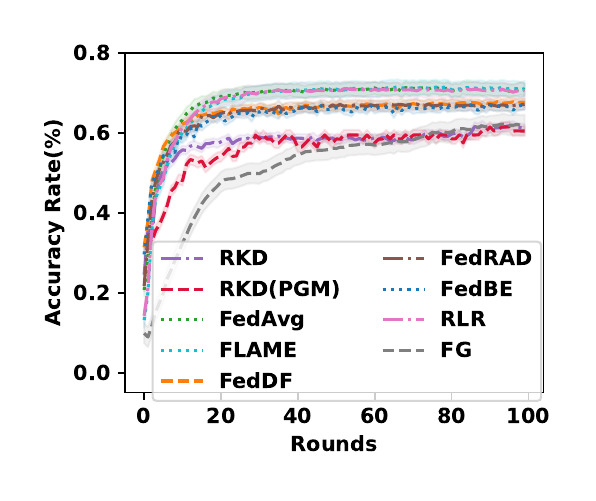}\label{fig:a4}}
{\includegraphics[width=0.325\columnwidth,height=2.5cm]{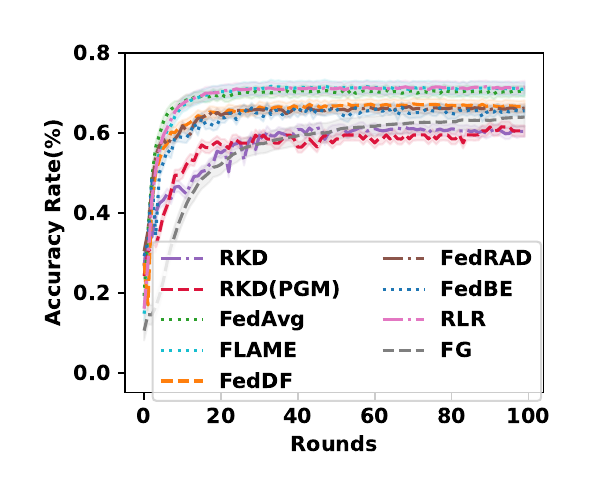}}
{\includegraphics[width=0.325\columnwidth,height=2.5cm]{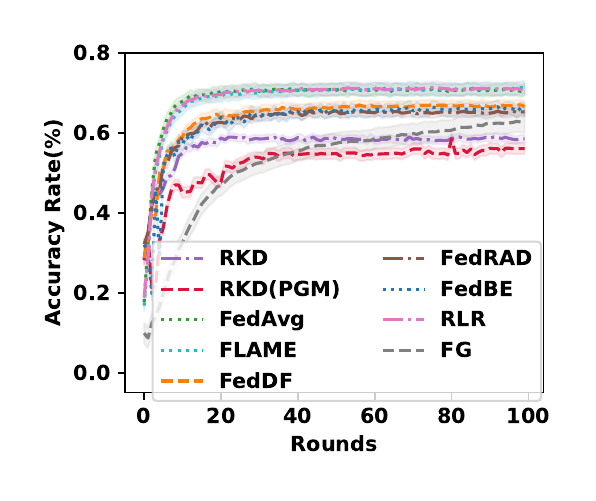}}\\
\subfloat[\scriptsize $20\%$ of A3FL]{\includegraphics[width=0.33\columnwidth,height=2.5cm]{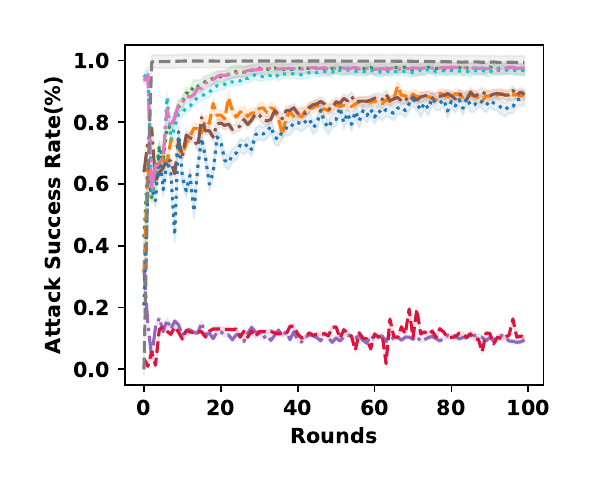}}
\subfloat[\scriptsize $40\%$ of A3FL]{\includegraphics[width=0.33\columnwidth,height=2.5cm]{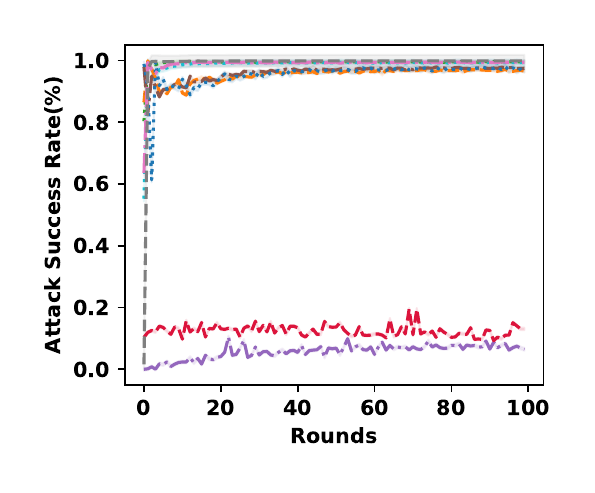}\label{fig:e4}}
\subfloat[\scriptsize $60\%$ of A3FL]{\includegraphics[width=0.33\columnwidth,height=2.5cm]{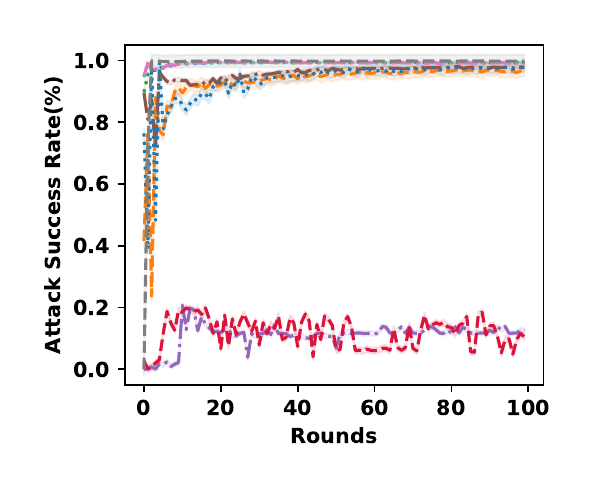}}

\caption{Performance of baselines and RKD on CIFAR-10 under Non-IID (\(\alpha=0.3\)), evaluated against $20\%$, $40\%$, and $60\%$ A3FL attacker clients.}
\label{CIFAR-A3FL}
\end{figure}

\begin{figure}[h!]
\centering
{\includegraphics[width=0.325\columnwidth,height=2.5cm]{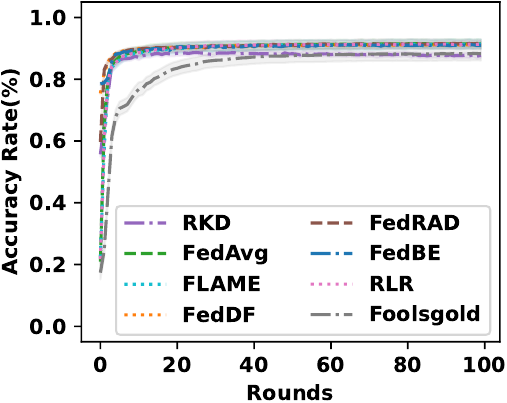}}
{\includegraphics[width=0.325\columnwidth,height=2.5cm]{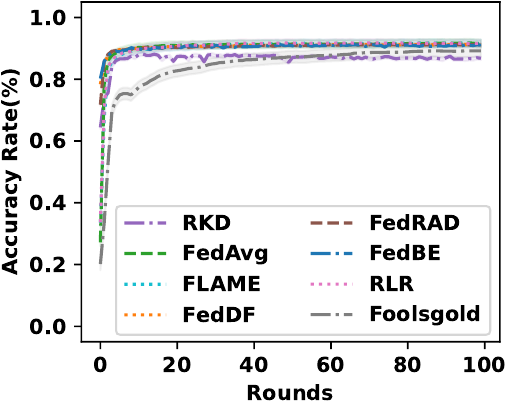}}
{\includegraphics[width=0.325\columnwidth,height=2.5cm]{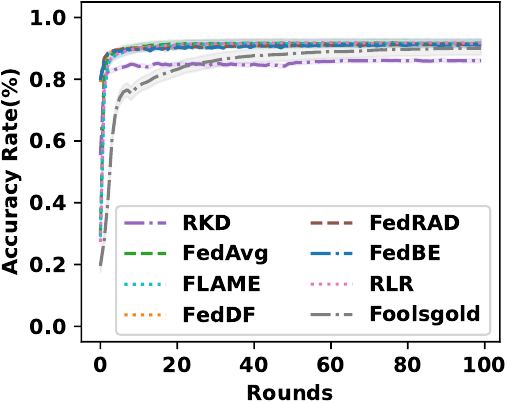}}\\
\subfloat[\scriptsize $20\%$ of A3FL]{\includegraphics[width=0.33\columnwidth,height=2.5cm]{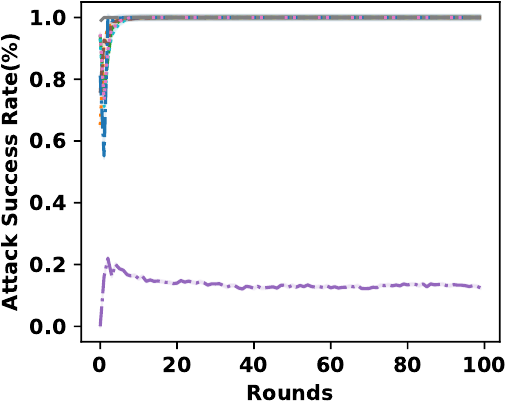}}
\subfloat[\scriptsize $40\%$ of A3FL]{\includegraphics[width=0.33\columnwidth,height=2.5cm]{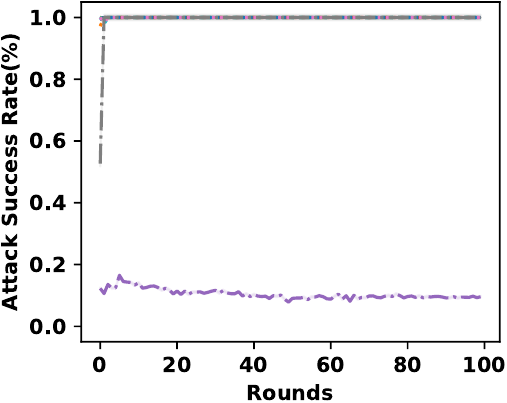}}
\subfloat[\scriptsize $60\%$ of A3FL]{\includegraphics[width=0.33\columnwidth,height=2.5cm]{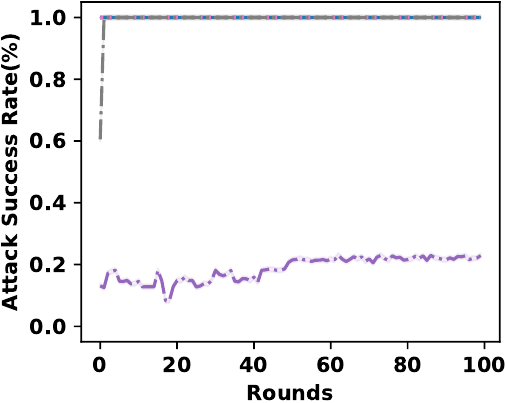}}

\caption{Performance of baselines and RKD on Fashion-MNIST under Non-IID (\(\alpha=0.3\)), evaluated against $20\%$, $40\%$, and $60\%$ A3FL attacker clients.}
\label{FashionMNIST-A3FL}
\end{figure}

\subsubsection{Defence Against A3FL Attack}

Under highly heterogeneous Non-IID conditions (\( \alpha = 0.3 \)), RKD demonstrated significant resilience against the A3FL attack on the CIFAR-10 and Fashion-MNIST datasets. As shown in Figures~\ref{CIFAR-A3FL} and~\ref{FashionMNIST-A3FL}, RKD achieved a substantially lower attack success rate while maintaining high accuracy compared to baseline methods.

These results confirm that RKD effectively distinguishes malicious from benign client updates and aggregates only reliable models. By restricting the dissemination of the updated global model to clients identified as benign, RKD prevents adversaries from adapting their strategies based on the latest global updates. In our primary approach—the \textbf{Exclusion Strategy}—malicious clients continue training with their current local models. Alternatively, the \textbf{Perturbation Strategy} (denoted as \(\text{RKD (PGM)}\)) provides suspected malicious clients with a minimally perturbed global model:
\(
M_{\text{pert}}^{r+1} = M_{\text{global}}^{r+1} + \eta,
\)
where \(\|\eta\| \approx 1 \times 10^{-4}\). This slight perturbation effectively obscures the precise state of the global model, thereby limiting the opportunity for adaptive adversaries to refine their attacks.

Comparative experiments show that \(\text{RKD (PGM)}\) maintains an average accuracy nearly identical to that of RKD using the Exclusion Strategy, while still mitigating adaptive attack risks. Overall, the experimental findings illustrate that RKD robustly mitigates backdoor attacks under Non-IID conditions.

\subsubsection{\textbf{Defense Against F3BA Attack}}

\begin{figure}[h!]
\centering
{\includegraphics[width=0.325\columnwidth,height=2.5cm]{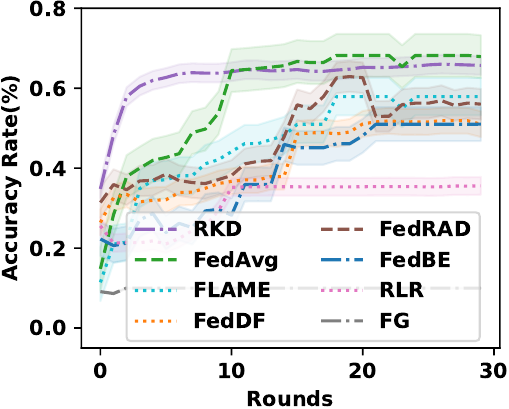}\label{fig:a7}}\
{\includegraphics[width=0.325\columnwidth,height=2.5cm]{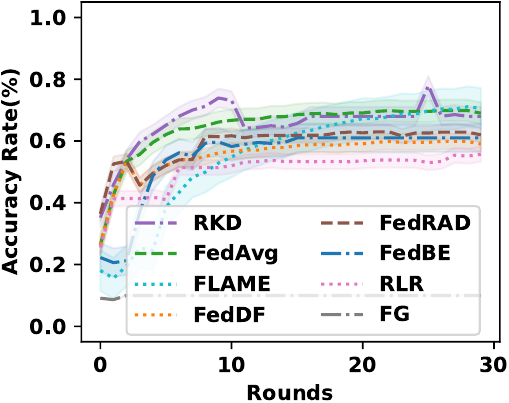}}
{\includegraphics[width=0.325\columnwidth,height=2.5cm]{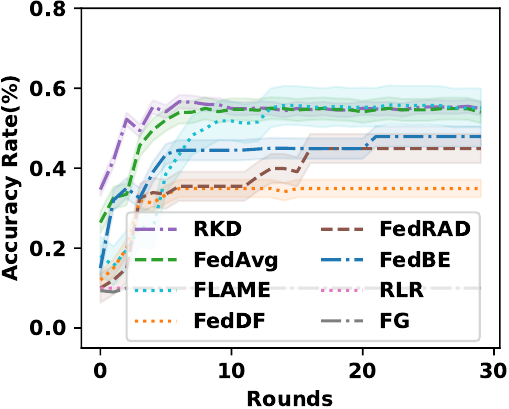}}\\
\subfloat[\scriptsize $20\%$ of F3BA]{\includegraphics[width=0.33\columnwidth,height=2.5cm]{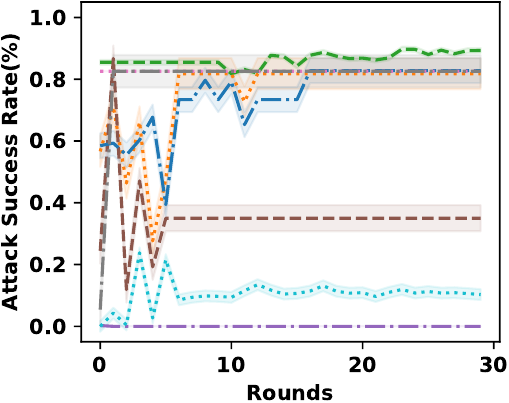}}
\subfloat[\scriptsize $40\%$ of F3BA]{\includegraphics[width=0.33\columnwidth,height=2.5cm]{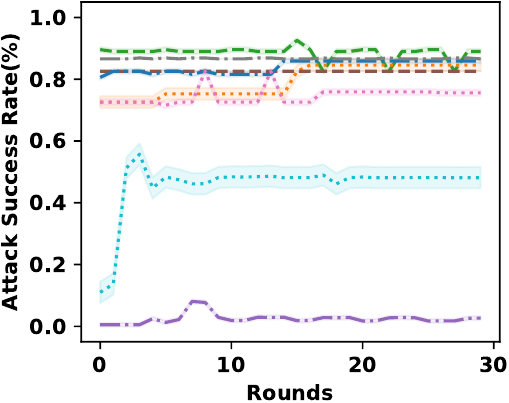}\label{fig:e7}}
\subfloat[\scriptsize $60\%$ of F3BA]{\includegraphics[width=0.33\columnwidth,height=2.5cm]{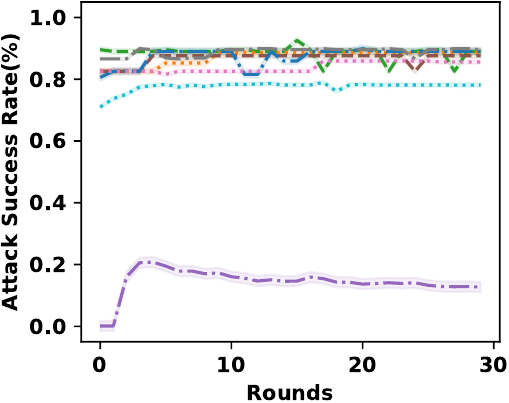}}

\caption{Performance of baselines and RKD on CIFAR-10 under Non-IID (\(\alpha=0.5\)), evaluated against $20\%$, $40\%$, and $60\%$ F3BA attacker clients.}
\label{CIFAR-F3BA}
\end{figure}

\begin{figure}[h!]
{\includegraphics[width=0.32\columnwidth,height=2.5cm]{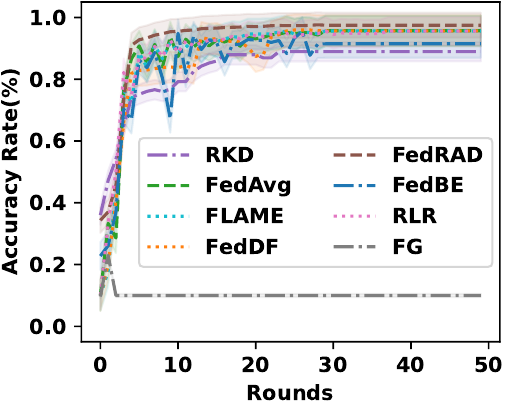}\label{fig:a5}}
{\includegraphics[width=0.32\columnwidth,height=2.5cm]{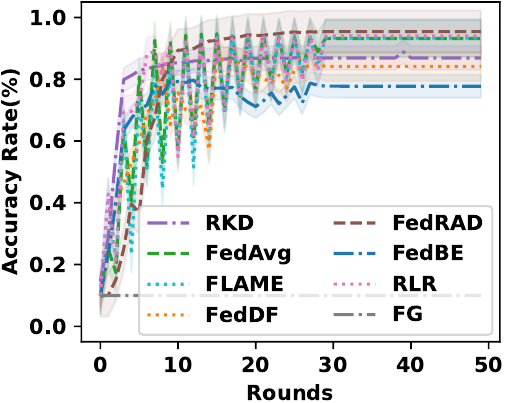}\label{fig:b5}}
{\includegraphics[width=0.32\columnwidth,height=2.5cm]{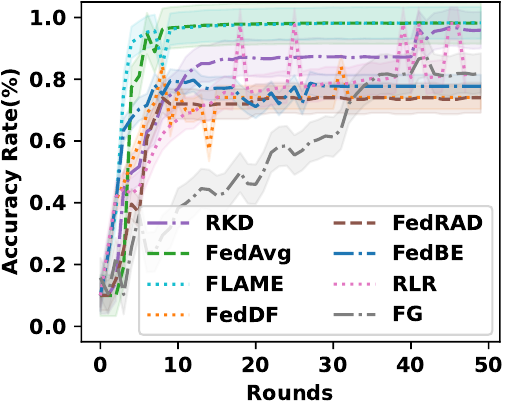}\label{fig:c5}}\\
\subfloat[\scriptsize $20\%$ of F3BA]{\includegraphics[width=0.32\columnwidth,height=2.5cm]{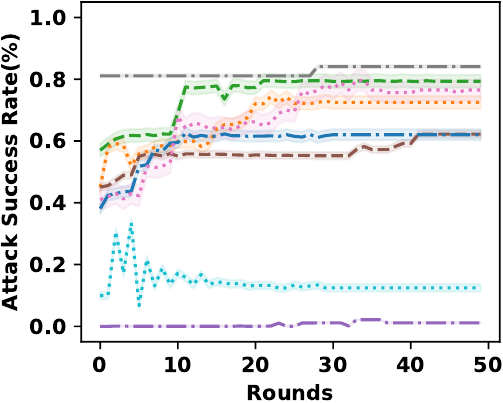}}
\subfloat[\scriptsize $40\%$ of F3BA]{\includegraphics[width=0.32\columnwidth,height=2.5cm]{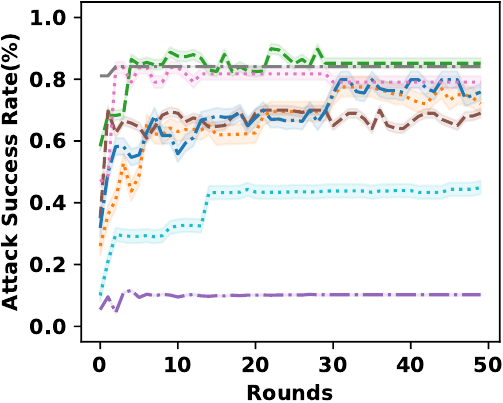}}
\subfloat[\scriptsize $60\%$ of F3BA]{\includegraphics[width=0.32\columnwidth,height=2.4cm]{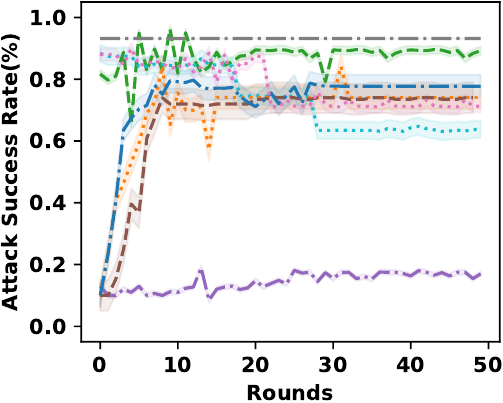}}

\caption{Performance of baselines and RKD on EMNIST under Non-IID (\(\alpha=0.5\)), evaluated against $20\%$, $40\%$, and $60\%$ F3BA attacker clients.}
\label{EMNIST-F3BA}
\end{figure}

RKD effectively defends against F3BA on CIFAR-10 and EMNIST datasets under non-IID conditions (\( \alpha = 0.5 \)), as substantiated by Figures \ref{CIFAR-F3BA} and \ref{EMNIST-F3BA}. Using cosine similarity-based clustering, RKD filters out anomalies from compromised clients and integrates knowledge distillation to maintain low ASR and high accuracy.

RKD's iterative training enhances the global model's accuracy and resilience, demonstrating its superiority over methods like FedAvg. This is particularly evident in high attacker ratios of \( 40\% \) and \( 60\% \), highlighting RKD's robust defence against sophisticated attacks like F3BA.

\subsubsection{\textbf{Defense Against DBA Attack}} 
\begin{figure}[h!]
{\includegraphics[width=0.32\columnwidth,height=2.5cm]{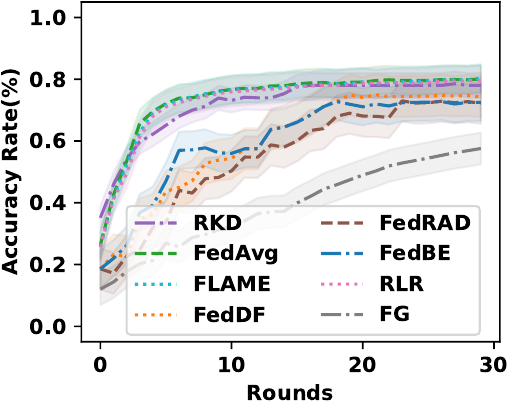}}
{\includegraphics[width=0.32\columnwidth,height=2.5cm]{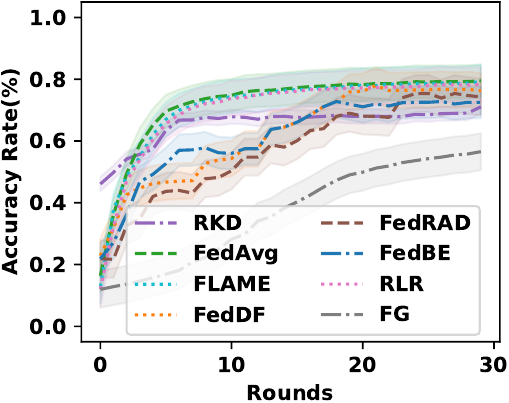}\label{fig:b4}}
{\includegraphics[width=0.32\columnwidth,height=2.5cm]{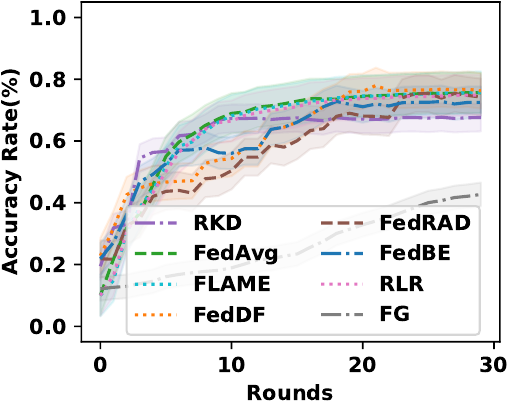}}\\
\subfloat[\scriptsize $20\%$ of DBA]{\includegraphics[width=0.32\columnwidth,height=2.5cm]{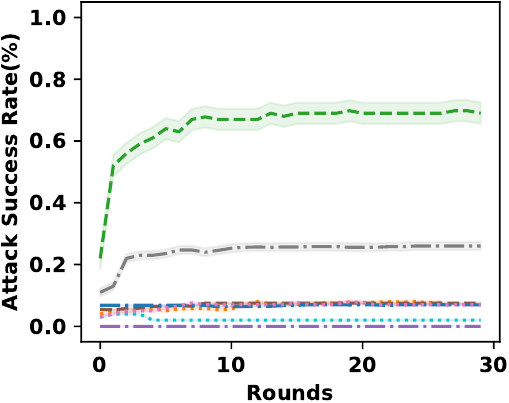}\label{fig:d4}}
\subfloat[\scriptsize $40\%$ of DBA]{\includegraphics[width=0.32\columnwidth,height=2.5cm]{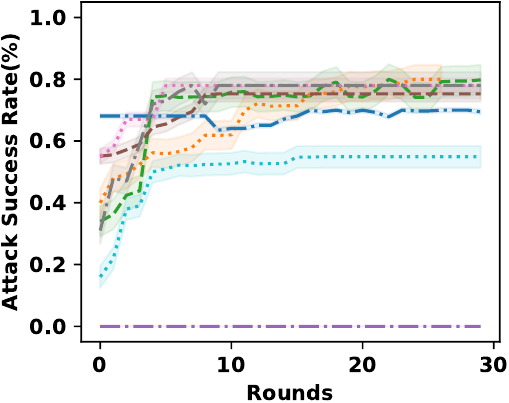}}
\subfloat[\scriptsize $60\%$ of DBA]{\includegraphics[width=0.32\columnwidth,height=2.5cm]{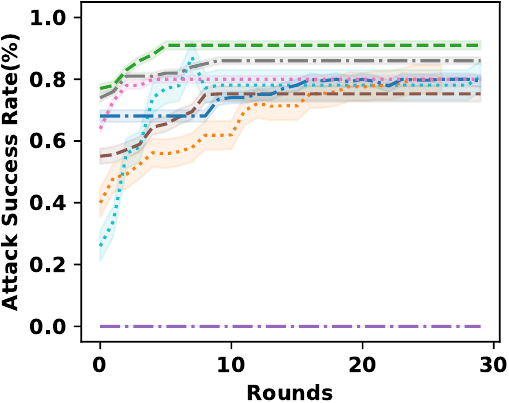}}

\caption{Performance of baselines and RKD on CIFAR-10 under Non-IID (\(\alpha=0.9\)), evaluated against $20\%$, $40\%$, and $60\%$ DBA attacker clients.}
\label{CIFAR-DBA}
\end{figure}

\begin{figure}[h!]
{\includegraphics[width=0.32\columnwidth,height=2.4cm]{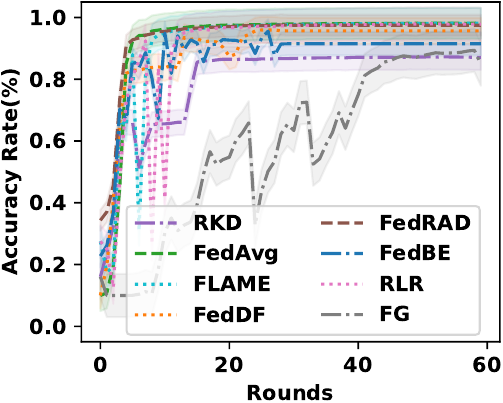}}
{\includegraphics[width=0.32\columnwidth,height=2.4cm]{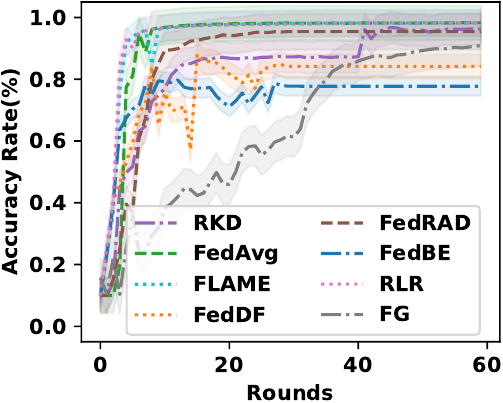}}
{\includegraphics[width=0.32\columnwidth,height=2.4cm]{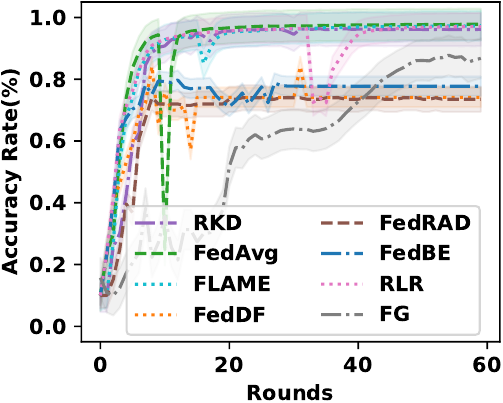}}\\
\subfloat[\scriptsize $20\%$ of DBA]{\includegraphics[width=0.32\columnwidth,height=2.4cm]{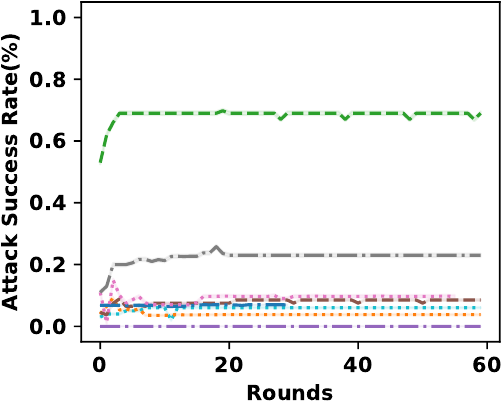}}
\subfloat[\scriptsize $40\%$ of DBA]{\includegraphics[width=0.32\columnwidth,height=2.4cm]{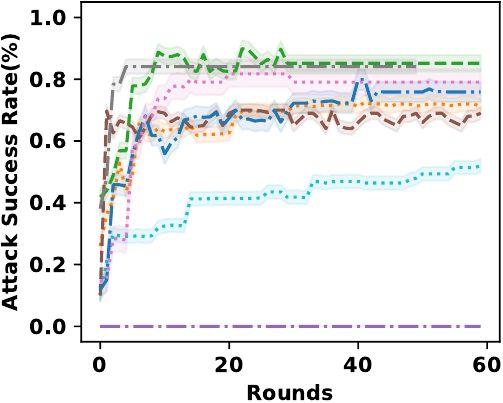}}
\subfloat[\scriptsize $60\%$ of DBA]{\includegraphics[width=0.32\columnwidth,height=2.4cm]{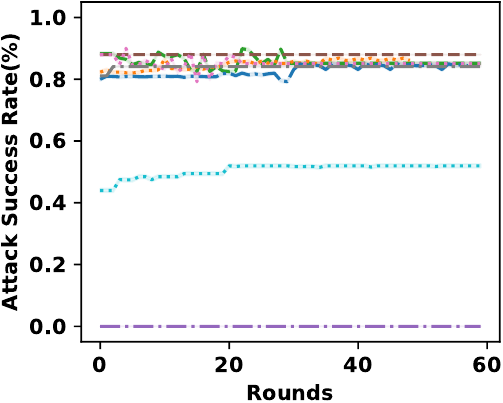}}

\caption{Performance of baselines and RKD on EMNIST under Non-IID (\(\alpha=0.9\)), evaluated against $20\%$, $40\%$, and $60\%$ DBA attacker clients.}
\label{EMNIST-DBA}
\end{figure}

RKD effectively defends against the Distributed Backdoor Attack (DBA) on CIFAR-10 and EMNIST under non-IID settings (\( \alpha = 0.9 \)), as shown in Figures \ref{CIFAR-DBA} and \ref{EMNIST-DBA}. Using cosine similarity-based clustering, RKD detects and isolates malicious updates. Median model selection ensures that only benign models contribute to the global model, minimizing subtle backdoor triggers.

During knowledge distillation, RKD synthesizes insights from vetted models into a robust aggregated model, enhancing generalizability and security. Compared to methods like FedDF, FedRAD, and FedBE, RKD provides superior protection by meticulously analyzing and distilling knowledge from selected models. This enables RKD to maintain high accuracy while significantly reducing the ASR, demonstrating its effectiveness against sophisticated attacks like DBA.

\subsubsection{Defense Against ADBA Attack}

\begin{figure}[h!]
{\includegraphics[width=0.32\columnwidth,height=2.7cm]{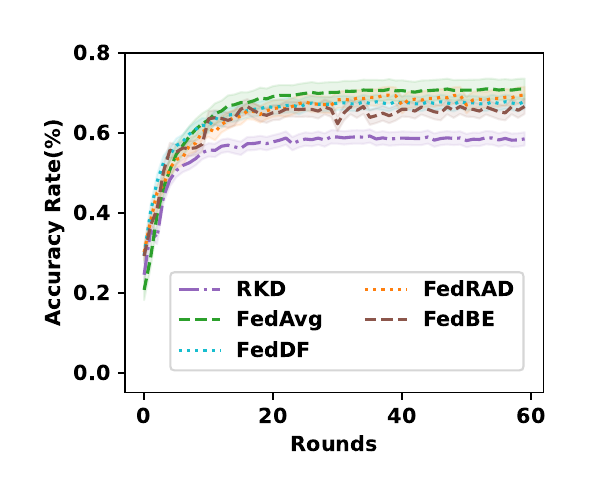}}
{\includegraphics[width=0.32\columnwidth,height=2.7cm]{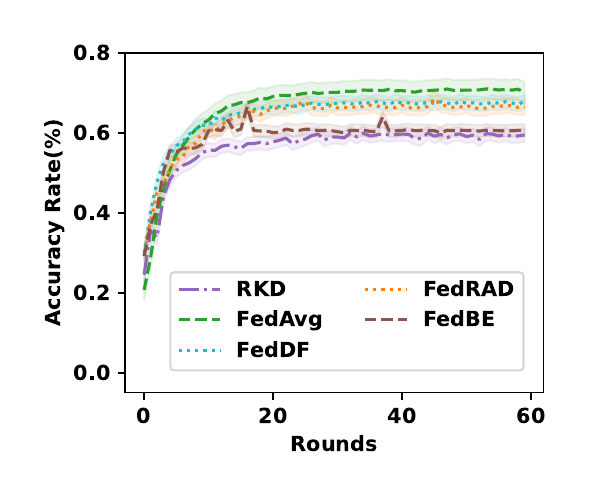}}
{\includegraphics[width=0.32\columnwidth,height=2.7cm]{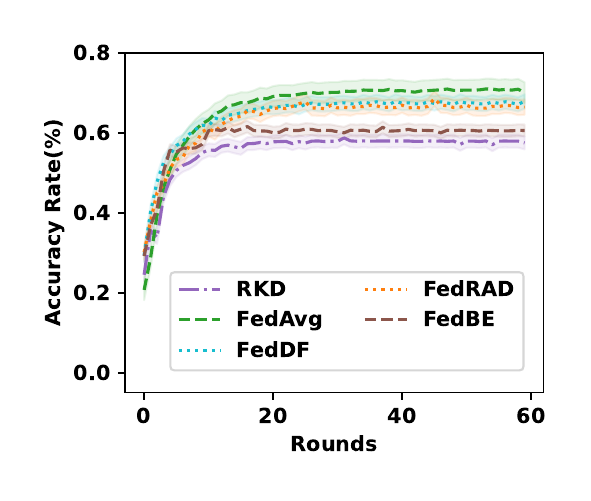}} \\
\subfloat[\scriptsize  $20\%$ of ADBA]{\includegraphics[width=0.32\columnwidth,height=2.7cm]{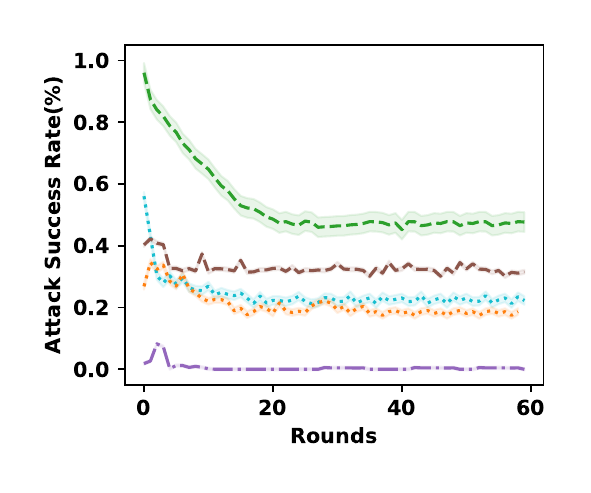}}
\subfloat[\scriptsize  $40\%$ of ADBA ]{\includegraphics[width=0.32\columnwidth,height=2.7cm]{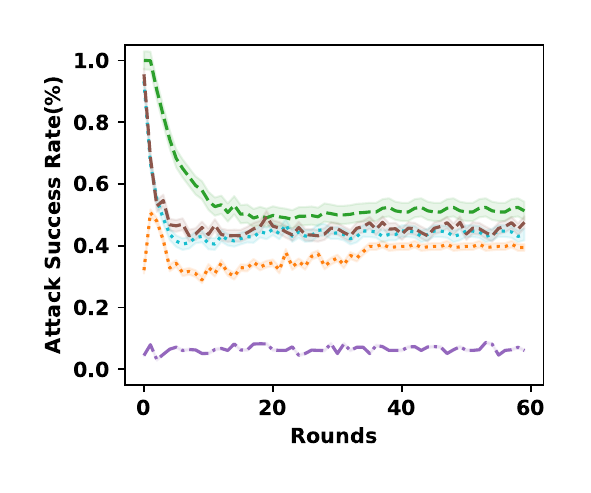}}
\subfloat[\scriptsize $60\%$ of ADBA]
{\includegraphics[width=0.32\columnwidth,height=2.7cm]{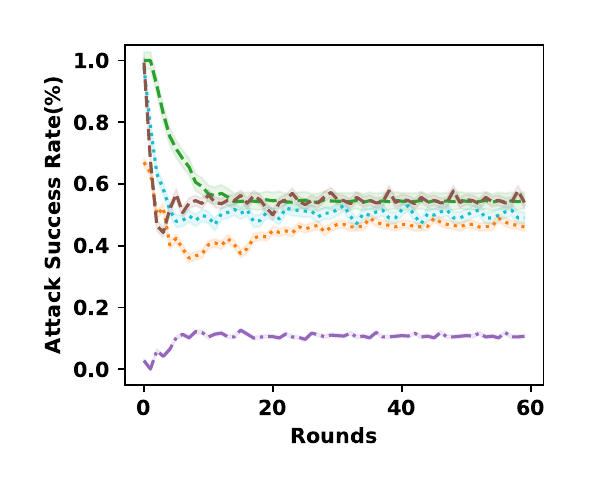}}

\caption{Performance of baselines and RKD on CIFAR-10 under Non-IID (\(\alpha=0.5\)), evaluated against $20\%$, $40\%$, and $60\%$ ADBA attacker clients.
}
\label{ADBAnonIID}
\end{figure}

The RKD framework robustly defends against Anti-Distillation Backdoor Attacks (ADBA) on CIFAR-10 under Non-IID conditions (\(\alpha = 0.5\)), as shown in Figure~\ref{ADBAnonIID}. Compared to FedAvg and other baseline methods, RKD effectively detects and mitigates ADBA backdoor attacks, demonstrating superior resilience and enhanced model integrity in challenging heterogeneous environments.

\subsubsection{\textbf{Defense Against TSBA Attack}} 
\begin{figure}[h!]

{\includegraphics[width=0.33\columnwidth,height=2.4cm]{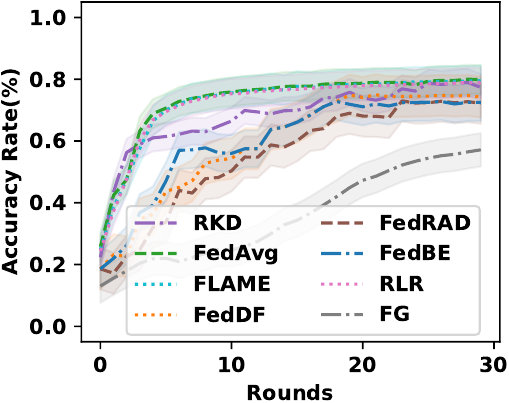}}
{\includegraphics[width=0.33\columnwidth,height=2.4cm]{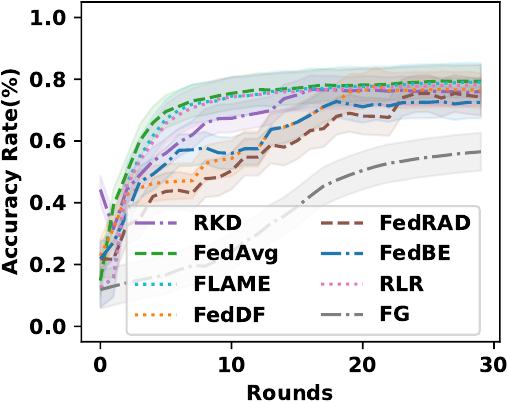}}
{\includegraphics[width=0.32\columnwidth,height=2.4cm]{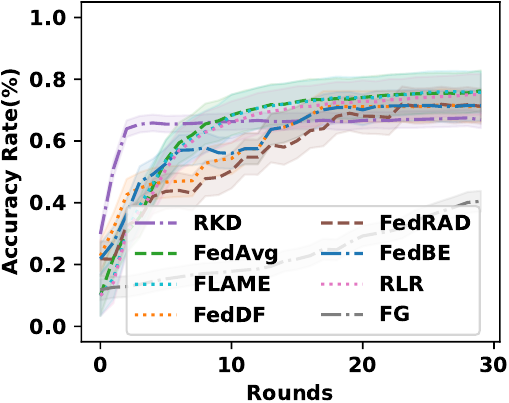}}\\
\subfloat[\scriptsize $20\%$ of TSBA]{\includegraphics[width=0.33\columnwidth,height=2.4cm]{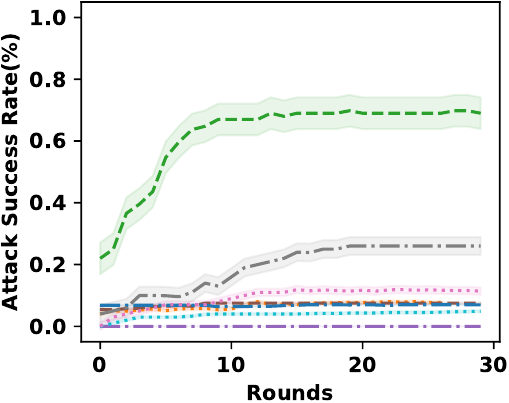}}
\subfloat[\scriptsize $40\%$ of TSBA]{\includegraphics[width=0.33\columnwidth,height=2.4cm]{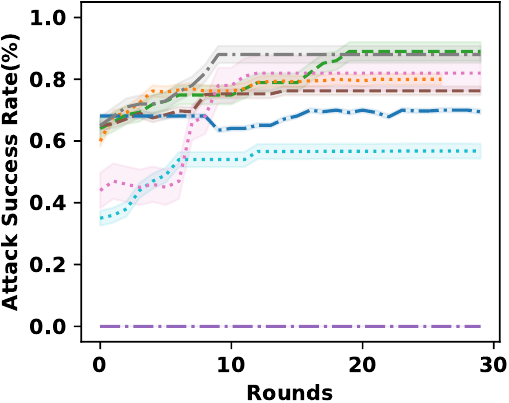}}
\subfloat[\scriptsize $60\%$ of TSBA]{\includegraphics[width=0.33\columnwidth,height=2.4cm]{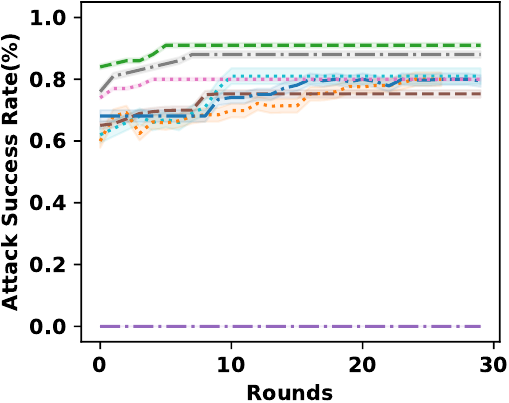}}

\caption{Performance of baselines and RKD on CIFAR-10 under Non-IID (\(\alpha=0.5\)), evaluated against $20\%$, $40\%$, and $60\%$ TSBA attacker clients.
}\label{CIFARTSBA}
\end{figure}

\begin{figure}[h!]
{\includegraphics[width=0.33\columnwidth,height=2.4cm]{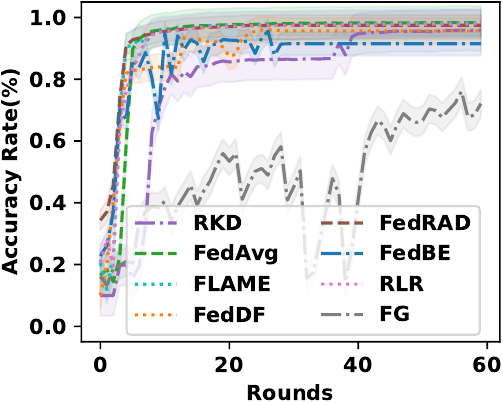}}
{\includegraphics[width=0.33\columnwidth,height=2.4cm]{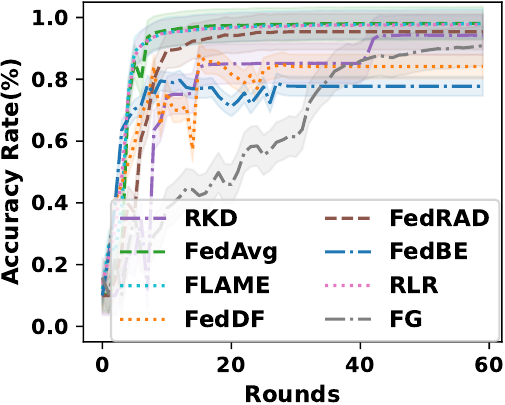}}
{\includegraphics[width=0.32\columnwidth,height=2.4cm]{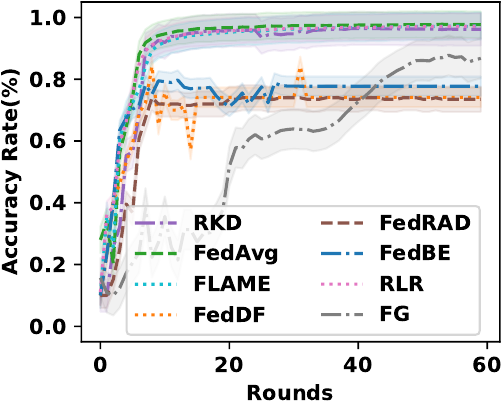}}\\
\subfloat[\scriptsize $20\%$ of TSBA]{\includegraphics[width=0.33\columnwidth,height=2.4cm]{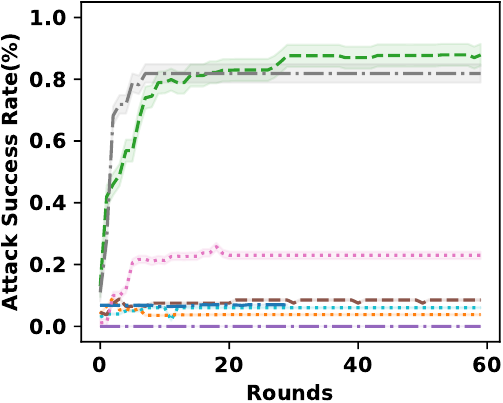}\label{fig:a2}}
\subfloat[\scriptsize $40\%$ of TSBA]{\includegraphics[width=0.33\columnwidth,height=2.4cm]{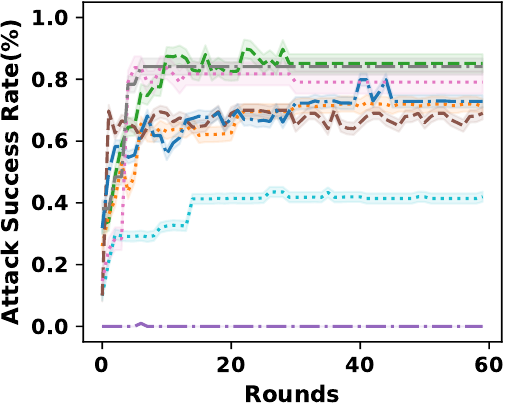}}
\subfloat[\scriptsize $60\%$ of TSBA]{\includegraphics[width=0.33\columnwidth,height=2.4cm]{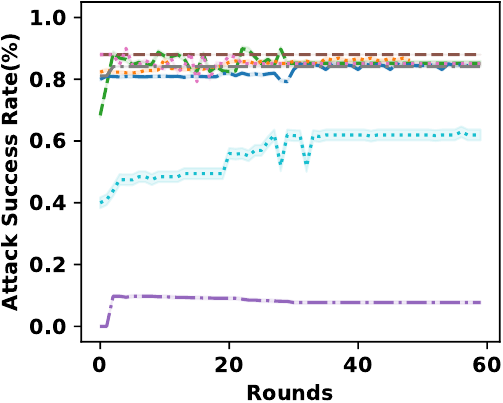}}

\caption{Performance of baselines and RKD on EMNIST under Non-IID (\(\alpha=0.5\)), evaluated against $20\%$, $40\%$, and $60\%$ TSBA attacker clients.
}\label{Emnist-TSBA}
\end{figure}

The RKD framework robustly defends against TSBA on CIFAR-10 and EMNIST under Non-IID conditions (\(\alpha = 0.5\)), as shown in Figures \ref{CIFARTSBA} and \ref{Emnist-TSBA}. 
RKD detects and mitigates TSBA manipulations, maintaining high accuracy and low ASR even with increased attacker ratios. Unlike other methods that falter under poisoned conditions, RKD excels with clean and poisoned datasets.

\subsubsection{The Impact of Heterogeneous Degree} 
\begin{figure}[h!]
{\includegraphics[width=0.32\columnwidth,height=2.4cm]{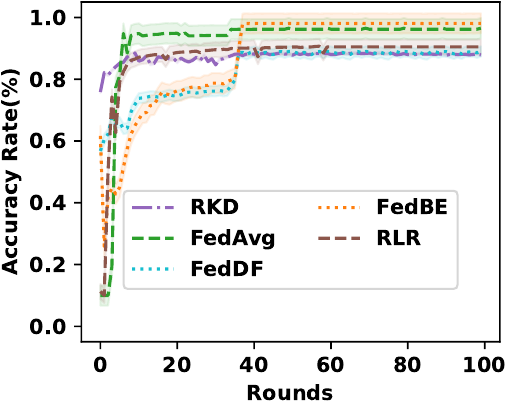}}
{\includegraphics[width=0.32\columnwidth,height=2.4cm]{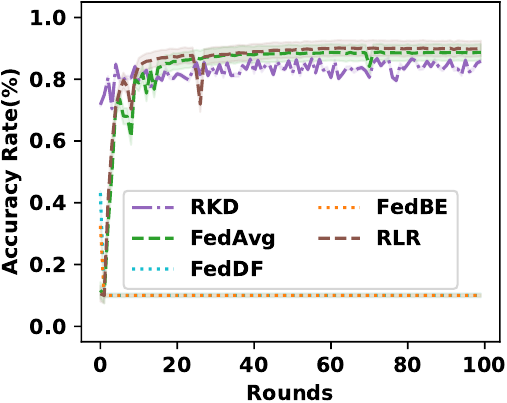}}
{\includegraphics[width=0.32\columnwidth,height=2.4cm]{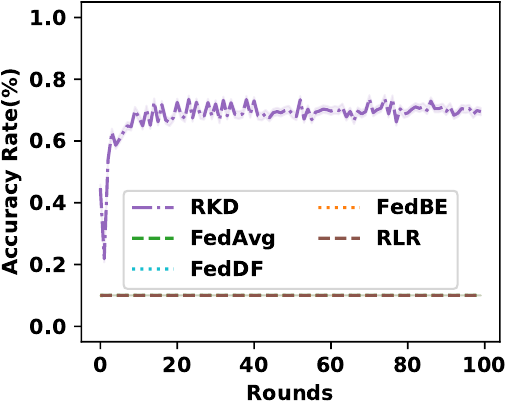}} \\
\subfloat[\scriptsize  Non-IID=$0.7$]{\includegraphics[width=0.33\columnwidth,height=2.4cm]{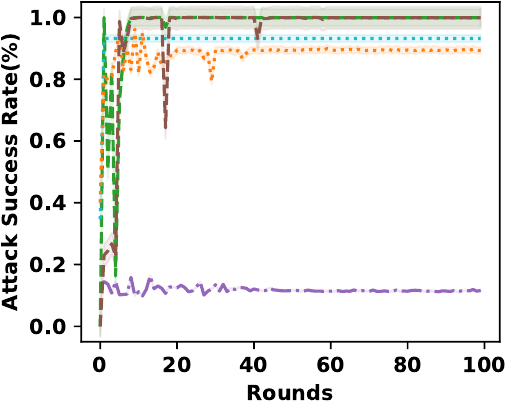}}
\subfloat[\scriptsize  Non-IID=$0.3$ ]{\includegraphics[width=0.33\columnwidth,height=2.4cm]{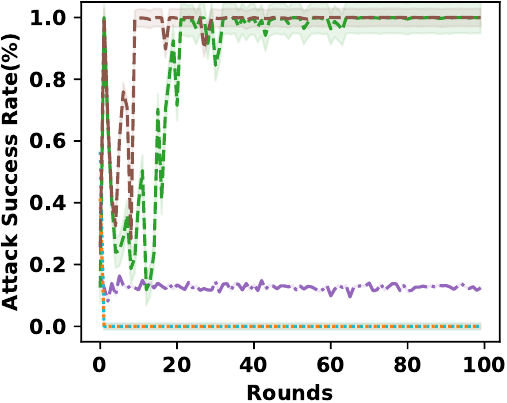}}
\subfloat[\scriptsize Non-IID=$0.1$]
{\includegraphics[width=0.33\columnwidth,height=2.4cm]{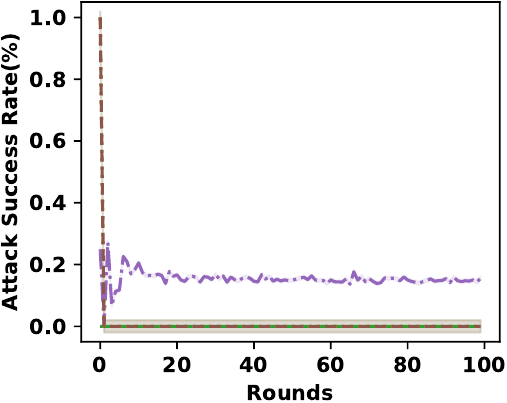}}

\caption{Performance impact of heterogeneous degrees on baselines and RKD on Fashion-MNIST, evaluated against $60\%$ F3BA attacker clients.
}
\label{fashion-nonIID}
\end{figure}

We evaluated baseline defence methods and the RKD framework under varying degrees of data heterogeneity, including moderate (\(\alpha = 0.7\)) and extreme (\(\alpha = 0.3, 0.1\)) Non-IID conditions (see Figure~\ref{fashion-nonIID}). Under extreme heterogeneity, many baseline methods achieve high accuracy on clean inputs but struggle to detect subtle backdoor triggers—resulting in elevated ASR. In contrast, RKD consistently maintains robust defence by effectively excluding malicious updates, which helps to suppress ASR while sustaining high MTA.

Notably, although extreme heterogeneity adversely impacts overall accuracy for all methods, RKD outperforms baseline defences by achieving a better balance between low ASR and high MTA. This indicates that a key contribution of our work is enhancing robustness in highly Non-IID scenarios. Moreover, under extremely Non-IID conditions (\(\alpha = 0.1\)), baseline methods often struggle to generalize, resulting in model collapse that leads to a low ASR—since their offline behavior prevents an accurate assessment of robustness. In contrast, RKD sustains stable learning and robust defence, achieving both high accuracy and a genuinely low ASR.

Overall, these results highlight RKD’s superior effectiveness in challenging heterogeneous data environments.

\subsubsection{Empirical Analysis of \( Q \) Sensitivity}

\begin{figure}[h!]
    \centering
    \includegraphics[width=0.8\linewidth]{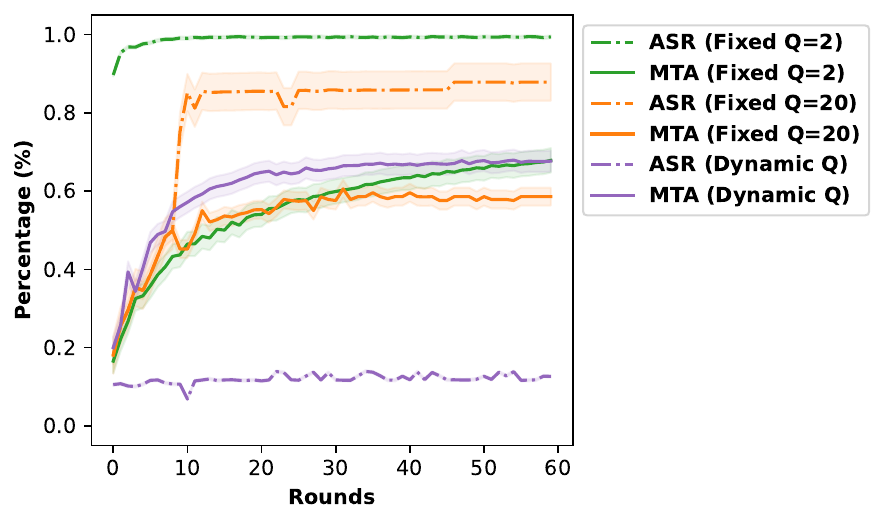}
    \caption{Impact of \( Q \) on MTA and ASR.}
    \label{fig:q_analysis}
\end{figure}

We evaluated the impact of the minimum cluster size \( Q \) on Main Task Accuracy (MTA) and Attack Success Rate (ASR) using the CIFAR-10 dataset under a Non-IID setting with 30 clients, 40\% of which were malicious and executing A3FL backdoor attacks. Figure~\ref{fig:q_analysis} presents the results.

When \( Q \) is fixed at 2, the resulting small clusters allow malicious updates to dominate, leading to a high ASR despite a relatively high MTA. In contrast, fixing \( Q \) at 20 excludes many benign updates, yielding slightly lower accuracy compared to the settings with \( Q = 2 \) or a dynamically adjusted \( Q \), while still exhibiting a higher ASR due to the misclassification of some malicious updates.

A dynamic adjustment of \( Q \) effectively mitigates these issues by excluding malicious updates while retaining the majority of benign ones, thus ensuring consistently high MTA and low ASR. These findings underscore the importance of dynamically tuning \( Q \) to reduce the influence of residual outliers and adversarial updates, thereby preserving the overall robustness and performance of the global model.

\subsubsection{Scalability Analysis} RKD enhances scalability by applying cosine similarity to model updates before clustering, transforming high-dimensional parameter vectors into scalar similarity scores. This dimensionality reduction significantly lowers computational complexity, making the clustering process more efficient. By avoiding clustering in the high-dimensional parameter space, RKD reduces both the time and resources required for defence operations. As shown in Table~\ref{tab:Defence_cpu}, RKD's defence time is $42.029$ seconds, substantially faster than FedDF and FedBE, which require $141.714$ and $198.765$ seconds, respectively. While RLR exhibits the shortest defence time, it compromises on detection accuracy due to its simplistic approach. FLAME is slightly more efficient in defence time, but RKD achieves a better balance between performance and robustness. These results highlight RKD's overall efficiency and scalability, demonstrating that its methodological design—specifically the use of cosine similarity and efficient clustering—provides robust defence without incurring significant computational overhead.

\begin{table}[h!]
\centering
\begin{tabular}{|l|r|}
\hline
\textbf{Method} & \textbf{Defense Time (seconds)} \\
\hline
RLR   & 0.020 \\
FLAME & 37.064 \\
RKD   & 42.029 \\
FedDF & 141.714 \\
FedBE & 198.765 \\
\hline
\end{tabular}
\caption{Comparative Defense Times}
\label{tab:Defence_cpu}
\end{table}

\subsection{Ablation Study}

\begin{figure}[htp]
{\includegraphics[width=0.32\columnwidth,height=2.3cm]{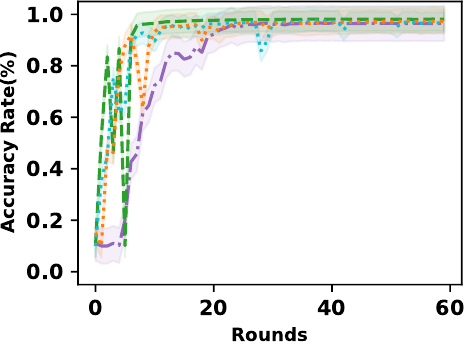}}
{\includegraphics[width=0.32\columnwidth,height=2.3cm]{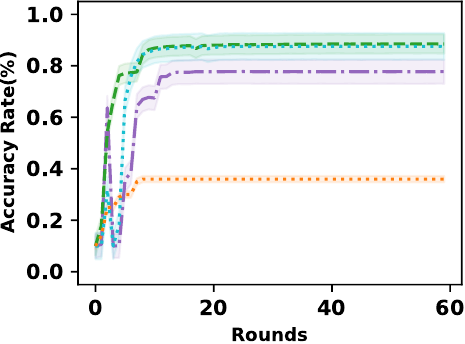}}
{\includegraphics[width=0.32\columnwidth,height=2.3cm]{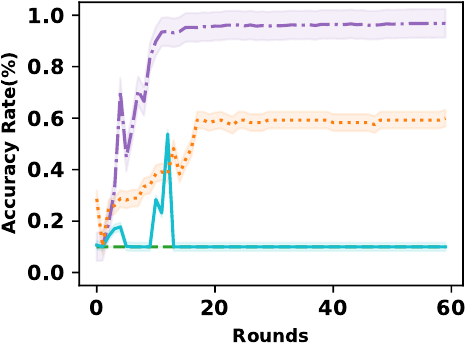}}

\subfloat[\scriptsize $20\%$ of DBA]{\includegraphics[width=0.32\columnwidth,height=2.3cm]{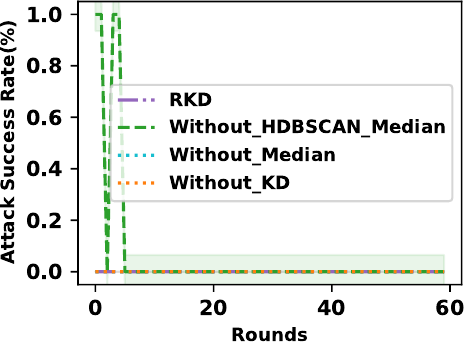}}
\subfloat[\scriptsize $40\%$ of DBA]{\includegraphics[width=0.32\columnwidth,height=2.3cm]{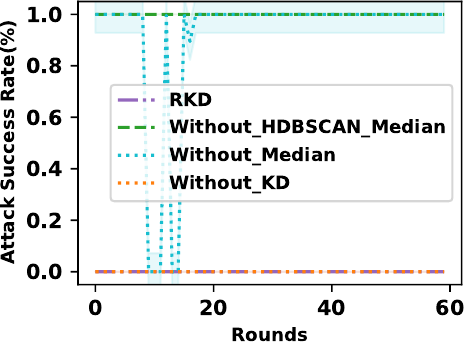}}
\subfloat[\scriptsize $60\%$ of F3BA]{\includegraphics[width=0.32\columnwidth,height=2.3cm]{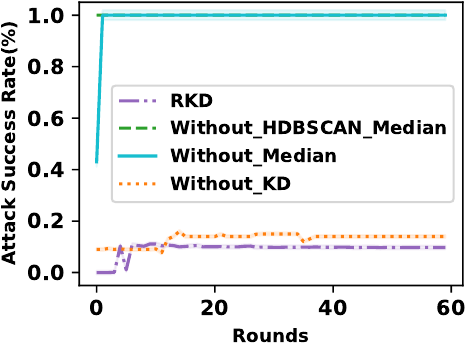}}
\caption{Ablation study of RKD method against  attacks.}\label{EMNIST0.5Ablation}
\end{figure}

We conducted ablation studies to evaluate the effectiveness of each component within the RKD framework against sophisticated backdoor attacks. Specifically, we analyzed the impact of removing key components: \textbf{Automated Clustering}, \textbf{Model Selection}, and \textbf{Knowledge Distillation}, as shown in Figure~\ref{EMNIST0.5Ablation}.

\textbf{RKD without Clustering and Model Selection (Median).} Removing both the \textit{Automated Clustering} and \textit{Model Selection} components, and leaving only the knowledge distillation process, significantly weakens the framework's defenses. While it performs reasonably well under a $20\%$ DBA attack, as the proportion of adversarial clients rises to $40\%$, its defensive capabilities drop sharply. At a $60\%$ F3BA attacker ratio, the model suffers from severe performance degradation, with misclassifications aligning with attackers' objectives. This highlights that without clustering, RKD is unable to effectively identify and isolate malicious updates, allowing backdoor attacks to compromise the global model.

\textbf{RKD without Model Selection (Median).} Excluding the \textit{ model selection} component while retaining HDBSCAN clustering, this variant effectively manages a $20\%$ DBA attack ratio. However, with a $40\%$ attack ratio, performance is noticeably decreased. Under a $60\%$ F3BA attacker ratio, the removal of model selection further impairs RKD's defense, resulting in increased ASR and reduced MTA. Even after the clustering phase effectively isolates most malicious updates, some malicious or anomalous updates may still be present. Without employing the median to mitigate the influence of residual outliers, these outliers can disproportionately affect the aggregated model, resulting in unstable ASR measurements. These findings underscore the critical role of model selection in refining the aggregation process by selecting models closest to the benign cluster centroid, thus enhancing robustness against higher proportions of attackers. 

\textbf{RKD without Knowledge Distillation.} Retaining both clustering and model selection but omitting the \textit{Knowledge Distillation} module, this variant successfully identifies and excludes malicious models, achieving a lower ASR. However, it suffers from a significant drop in accuracy, approximately $20\%$ lower than the full RKD framework. This underscores the crucial role of KD in transferring knowledge from the selected ensemble of models to the global model, which enhances both accuracy and generalization, particularly across Non-IID settings. Without KD, the framework is unable to effectively integrate and refine the knowledge from ensemble models, leading to poor performance in Non-IID environments despite successful isolation of malicious updates. KD specifically addresses the challenge of generalization in Non-IID data scenarios, ensuring the model remains robust and effective.

\section{Conclusion and Future Work}
We present RKD, a novel framework designed to mitigate sophisticated backdoor attacks in FL under Non-IID conditions. Our extensive evaluations on CIFAR-10, EMNIST, and Fashion-MNIST datasets demonstrated RKD's effectiveness in maintaining high accuracy levels while significantly reducing attack success rates. RKD's automated clustering, model selection, and knowledge distillation components collectively enhance the robustness and integrity of FL systems. Our results underscore RKD's superiority over existing defence methods, offering a robust solution to preserve the efficacy and security of FL in diverse and adversarial environments.

\section*{Acknowledgments}
We gratefully acknowledge the support of the Saudi Arabian Ministry of Education, the Saudi Arabian Cultural Bureau in London, and Umm Al-Qura University in Makkah. We are thankful to the High-End Computing facility at Lancaster University for providing essential computational resources. We also acknowledge the scholarship awarded by the 3rd IEEE Conference on Secure and Trustworthy Machine Learning, which facilitated our participation. Finally, we express our sincere gratitude to Dr. Matheus Aparecido do Carmo Alves for his invaluable review.

\appendix
\section{Additional Experiments under IID}
\subsection{Defense Against A3FL Attack Under IID Conditions.} 
\begin{figure}[htp]
{\includegraphics[width=0.32\columnwidth,height=2.4cm]{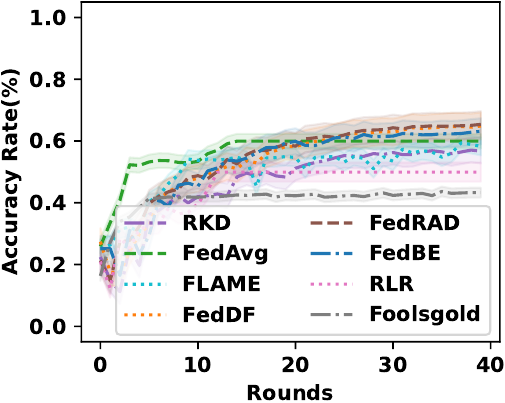}}
{\includegraphics[width=0.32\columnwidth,height=2.4cm]{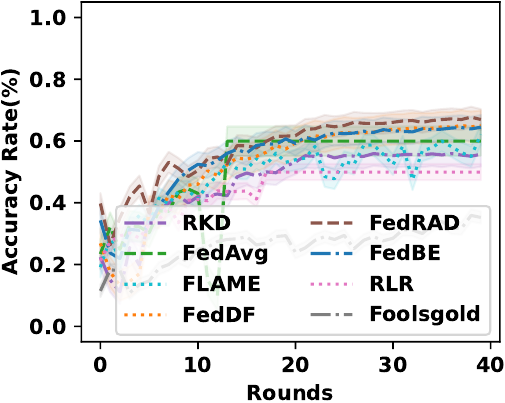}}
{\includegraphics[width=0.32\columnwidth,height=2.4cm]{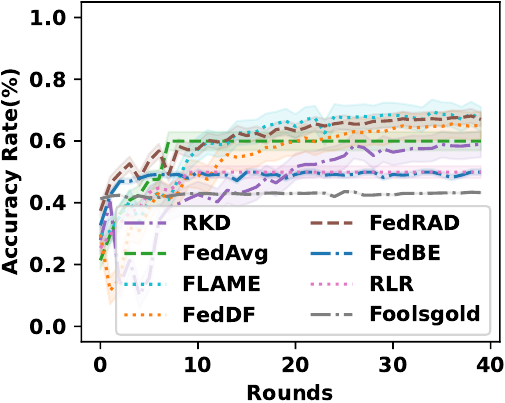}} \\
\subfloat[\scriptsize  $20\%$ of A3FL]{\includegraphics[width=0.32\columnwidth,height=2.4cm]{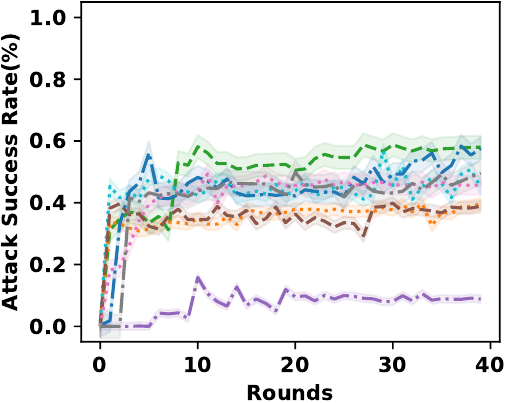}}
\subfloat[\scriptsize  $40\%$ of A3FL ]{\includegraphics[width=0.32\columnwidth,height=2.4cm]{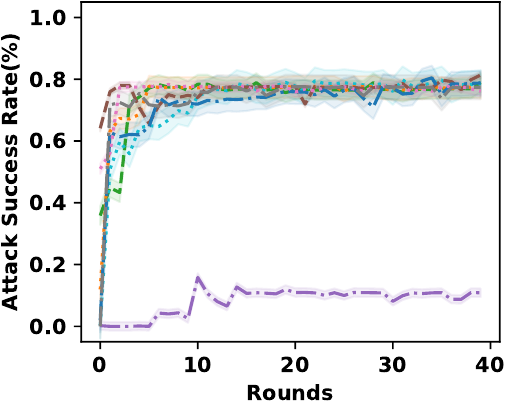}}
\subfloat[\scriptsize $60\%$ of A3FL]
{\includegraphics[width=0.32\columnwidth,height=2.4cm]{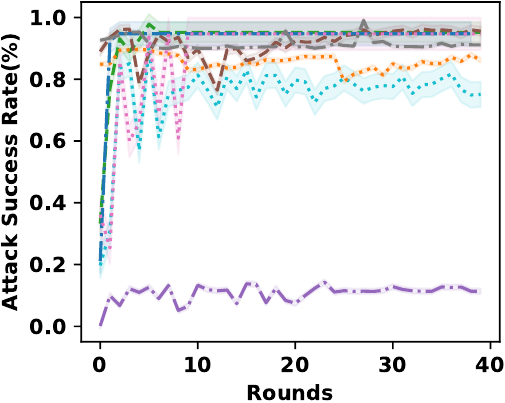}}

\caption{Performance of baselines and RKD on CIFAR-10 under IID settings against A3FL attacker clients.
}
\label{A3FLIID}
\end{figure}

As shown in Figure~\ref{A3FLIID}, RKD outperforms baseline models in defending against the A3FL attack on the CIFAR-10 dataset under IID conditions, across varying attacker ratios ($20\%$, $40\%$, and $60\%$). RKD's robustness against adversarial attacks in IID settings underscores its effectiveness as a defence mechanism. The baseline models also demonstrate some inherent robustness at lower attacker ratios ($20\%$ and $40\%$), which can be attributed to the uniform data distribution in IID settings that allows models to learn more generalizable patterns, making it harder for attacks to succeed.

\subsection{Defense Against F3BA Attack Under IID Conditions.} 
\begin{figure}[htp]
{\includegraphics[width=0.32\columnwidth,height=2.4cm]{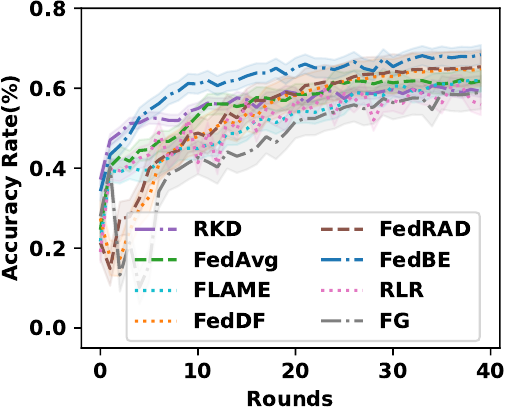}}
{\includegraphics[width=0.32\columnwidth,height=2.4cm]{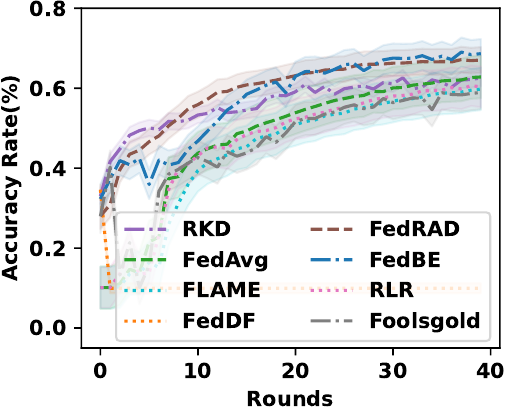}}
{\includegraphics[width=0.32\columnwidth,height=2.4cm]{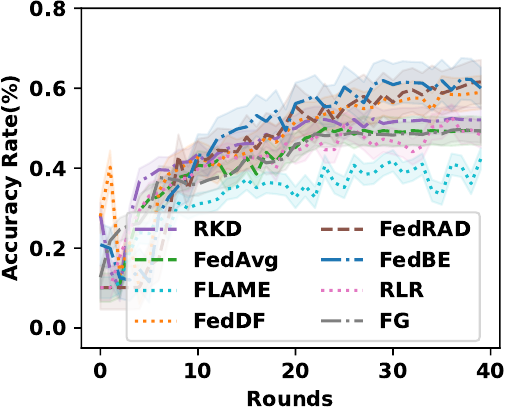}} \\
\subfloat[\scriptsize  $20\%$ of F3BA]{\includegraphics[width=0.32\columnwidth,height=2.4cm]{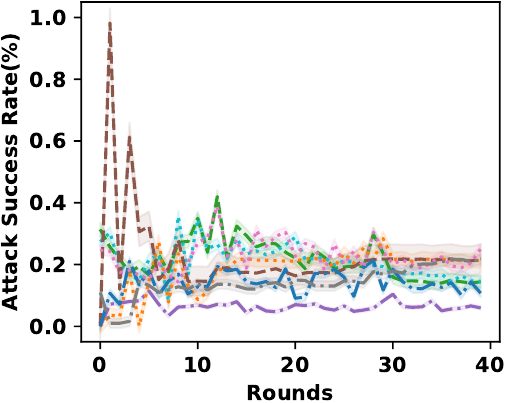}}
\subfloat[\scriptsize  $40\%$ of F3BA ]{\includegraphics[width=0.32\columnwidth,height=2.4cm]{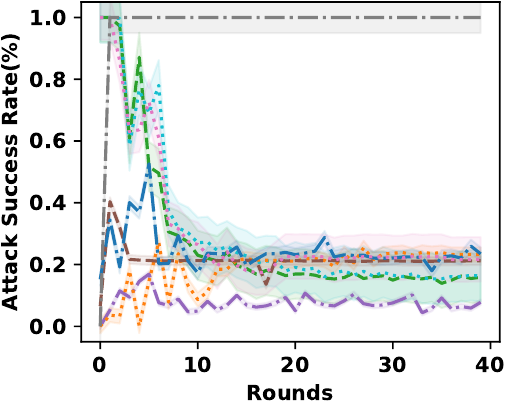}}
\subfloat[\scriptsize $60\%$ of F3BA]
{\includegraphics[width=0.32\columnwidth,height=2.4cm]{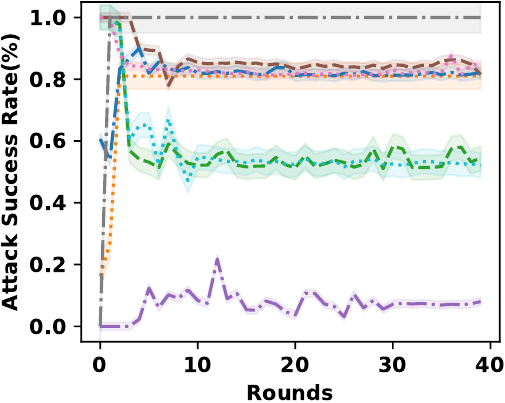}}

\caption{Performance of baselines and RKD on CIFAR-10 under IID settings against F3BA attacker clients.
}
\label{F3BAIID}
\end{figure}

Figure~\ref{F3BAIID} illustrates RKD's effective defense against the F3BA attack on the CIFAR-10 dataset under IID conditions. RKD consistently achieves lower ASR and higher MTA compared to baseline methods across different attacker ratios. This superior performance highlights RKD's potential as a reliable defense in adversarial federated learning environments. Baseline models show some robustness against $20\%$ of F3BA attackers, likely due to the IID settings enabling the models to learn generalizable patterns that are less susceptible to manipulation by adversarial clients.

\subsection {Performance Under Non-IID and IID Conditions Without Attacks}

\begin{figure}[htp]
\centering
{\includegraphics[width=0.33\columnwidth,height=2.6cm]{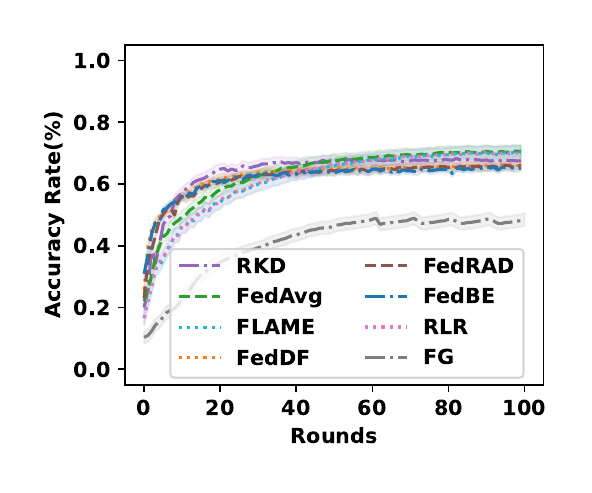}}
{\includegraphics[width=0.33\columnwidth,height=2.6cm]{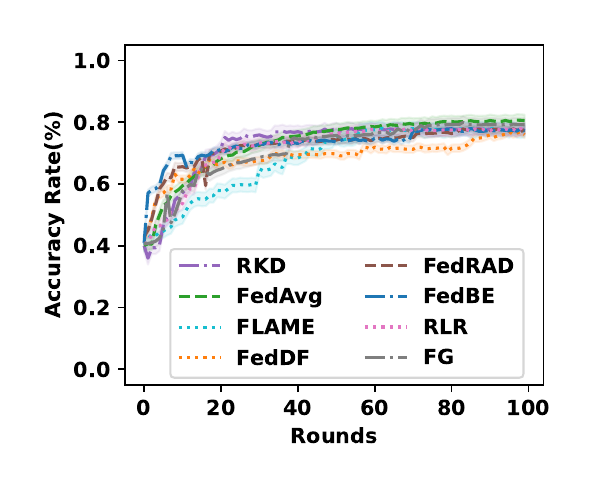}} \\
\subfloat[\scriptsize Non-IID]{\includegraphics[width=0.33\columnwidth,height=2.6cm]{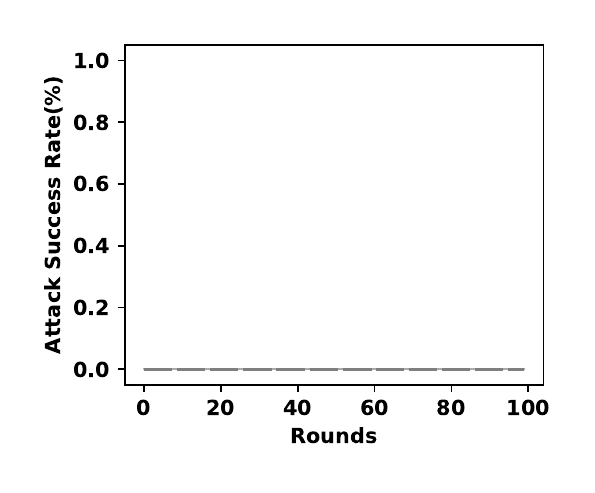}}
\subfloat[\scriptsize IID]{\includegraphics[width=0.33\columnwidth,height=2.6cm]{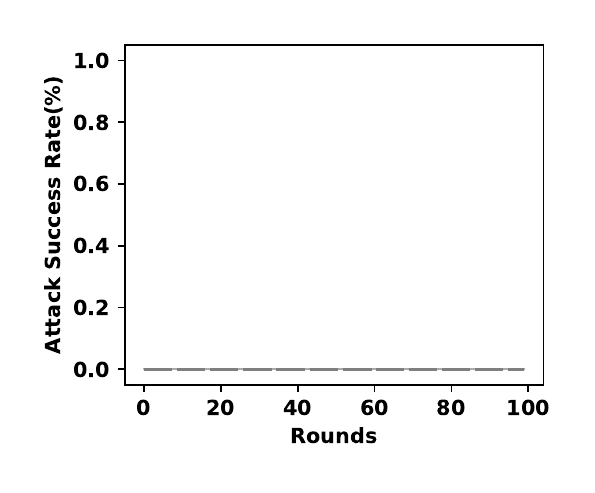}}

\caption{Performance of baselines and RKD on CIFAR-10 under Non-IID and IID conditions with no attack.}
\label{Noattack}
\end{figure}

We evaluated the performance of RKD and baseline methods under both Non-IID and IID data distributions in the absence of attacks. The results, illustrated in Figure~\ref{Noattack}, demonstrate that the attack success rate remains close to zero across all methods when no attacks are present. Additionally, models trained under IID conditions consistently achieve higher accuracy compared to those trained under Non-IID. 

This difference is attributed to the uniform distribution of data in IID settings, which facilitates more effective training and generalization. Conversely, the lower accuracy observed in Non-IID models reflects the challenges posed by data heterogeneity, highlighting the importance of robust training strategies to handle Non-IID data distributions effectively.

\end{document}